\documentclass[
aip,
rsi,
twocolumn,
reprint,
floatfix,
citeautoscript,
]{revtex4-2}

\usepackage[T1]{fontenc}
\usepackage[utf8]{inputenc}
\usepackage{newtxtext}
\usepackage{newtxmath}
\usepackage{chemformula}
\usepackage{siunitx}

\usepackage[
unicode,
colorlinks = true,
allcolors = blue,
]{hyperref}

\usepackage{cleveref}

\usepackage{microtype}

\usepackage{comment}

\usepackage{xcolor}

\usepackage{graphicx}

\usepackage{glossaries}
\glsdisablehyper

\newacronym[sort={b-NMR}]{bnmr}{\ensuremath{\beta}-NMR}{\ensuremath{\beta}-detected nuclear magnetic resonance}
\newacronym[sort={b-NQR}]{bnqr}{\ensuremath{\beta}-NQR}{\ensuremath{\beta}-detected nuclear quadrupole resonance}
\newacronym{nmr}{NMR}{nuclear magnetic resonance}
\newacronym{slr}{SLR}{spin-lattice relaxation}
\newacronym{musr}{\ensuremath{\mu}SR}{muon spin rotation}
\newacronym[sort={muSR}]{le-musr}{LE-\ensuremath{\mu}SR}{low-energy muon spin rotation}
\newacronym{mu}{\ensuremath{\mu^+}}{positive muon}
\newacronym{li-8}{\ch{^{8}Li^{+}}}{radioactive Li-8 ion}
\newacronym{li-8-neut}{\ch{^{8}Li}}{radioactive Li-8}
\newacronym{nb-93}{\ch{^{93}Nb}}{host Nb}
\newacronym{li-7}{\ch{^{7}Li^{+}}}{stable Li-7 ion}
\newacronym{c-12}{\ch{^{12}C^{+}}}{stable C-12 ion}
\newacronym{t-half}{\ensuremath{T_{1/2}}}{half-life}
\newacronym{t-lifetime}{\ensuremath{\tau}}{radioactive lifetime}

\newacronym{psi}{PSI}{Paul Scherrer Institute}
\newacronym{isac}{ISAC}{Isotope Separator and Accelerator}

\newacronym{rf}{RF}{radio frequency}
\newacronym{dc}{DC}{direct current}
\newacronym{em}{EM}{electromagnetic}
\newacronym{srf}{SRF}{superconducting radio frequency}
\newacronym{linac}{LINAC}{linear accelerator}
\newacronym{cw}{CW}{continuous wave}
\newacronym{rib}{RIB}{radioactive ion beam}

\newacronym{nb}{\ch{Nb}}{niobium}
\newacronym{tc}{\ensuremath{T_c}}{critical temperature}
\newacronym{b-surf}{\ensuremath{B_\text{surf}}}{surface magnetic field}
\newacronym{b-c1}{\ensuremath{B_{c1}}}{lower critical field}
\newacronym{b-c2}{\ensuremath{B_{c2}}}{upper critical field}
\newacronym{b-c}{\ensuremath{B_\text{crit}}}{thermodynamic critical field}
\newacronym{b-sh}{\ensuremath{B_{sh}}}{superheating field}
\newacronym{b-vp}{\ensuremath{B_\text{vp}}}{field of first vortex penetration}
\newacronym{lpd}{\ensuremath{\lambda_L}}{London penetration depth}
\newacronym{RRR}{RRR}{residual resistivity ratio}

\newacronym{uhv}{UHV}{ultra-high vacuum}
\newacronym{cf}{CF}{ConFlat}
\newacronym{hv}{HV}{high voltage}
\newacronym{lv}{LV}{low voltage}
\newacronym{LHe}{LHe}{liquid helium}
\newacronym{rga}{RGA}{residual gas analyzer}

\newacronym{lpm}{LPM}{linear position monitor}
\newacronym{fc}{FC}{Faraday cup}
\newacronym{ccd}{CCD}{charge-coupled device}

\newacronym{od}{OD}{outer diameter}
\newacronym{id}{ID}{inner diameter}
\newacronym{tig}{TIG}{tungsten inert gas}
\newacronym{iacs}{IACS}{the International Annealed Copper Standard}
\newacronym{ofhc}{OFHC}{oxygen-free high thermal conductivity}

\newacronym{gpm}{GPM}{gallons per minute}

\newacronym{epics}{EPICS}{Experimental Physics and Industrial Control System}
\newacronym{pid}{PID}{proportional–integral–derivative}

\newacronym{pmt}{PMT}{photomultiplier tube}
\newacronym{daq}{DAQ}{data aquisition}

\newacronym{wiener}{}{WIENER Plein \& Baus, Corp.}

\newacronym{gui}{GUI}{graphical user interface}
\newacronym{api}{API}{application programming interface}

\newacronym{fwhm}{FWHM}{full width at half maximum}

\newacronym{srim}{SRIM}{Stopping and Range of Ions in Matter}

\newacronym{fem}{FEM}{finite element method}

\newcommand{\LiEightPlus}{\ch{^{8}Li^{+}}}
\newcommand{\LiEight}{\ch{^{8}Li}}
\newcommand{\NbNinetyThree}{\ch{^{93}Nb}}
\newcommand{\LiSevenPlus}{\ch{^{7}Li^{+}}}
\newcommand{\CTwelvePlus}{\ch{^{12}C^{+}}}

\begin{document}
		\title{A New High Parallel-Field Spectrometer at TRIUMF's $\beta$-NMR Facility}
	
	\author{Edward~Thoeng}
	\email[E-mail: ]{ethoeng@triumf.ca}
	\affiliation{Department of Physics \& Astronomy, University of British Columbia, 6224 Agricultural Road, Vancouver, BC V6T~1Z1, Canada}
	\affiliation{TRIUMF, 4004 Wesbrook Mall, Vancouver, BC V6T~2A3, Canada}
	
	\author{Ryan~M.~L.~McFadden}
	\email[E-mail: ]{rmlm@triumf.ca}
	\affiliation{TRIUMF, 4004 Wesbrook Mall, Vancouver, BC V6T~2A3, Canada}
	\affiliation{Department of Physics and Astronomy, University of Victoria, 3800 Finnerty Road, Victoria, BC V8P~5C2, Canada}
	
	\author{Suresh~Saminathan}
	\affiliation{TRIUMF, 4004 Wesbrook Mall, Vancouver, BC V6T~2A3, Canada}
	
	\author{Gerald~D.~Morris}
	\affiliation{TRIUMF, 4004 Wesbrook Mall, Vancouver, BC V6T~2A3, Canada}
	
	\author{Philipp~Kolb}
	\affiliation{TRIUMF, 4004 Wesbrook Mall, Vancouver, BC V6T~2A3, Canada}
	
	\author{Ben~Matheson}
	\affiliation{TRIUMF, 4004 Wesbrook Mall, Vancouver, BC V6T~2A3, Canada}
	
	\author{Md~Asaduzzaman}
	\affiliation{TRIUMF, 4004 Wesbrook Mall, Vancouver, BC V6T~2A3, Canada}
	\affiliation{Department of Physics and Astronomy, University of Victoria, 3800 Finnerty Road, Victoria, BC V8P~5C2, Canada}
	
	\author{Richard~Baartman}
	\affiliation{TRIUMF, 4004 Wesbrook Mall, Vancouver, BC V6T~2A3, Canada}
	
	\author{Sarah~R.~Dunsiger}
	\affiliation{TRIUMF, 4004 Wesbrook Mall, Vancouver, BC V6T~2A3, Canada}
	\affiliation{Department of Physics, Simon Fraser University, 8888 University Drive, Burnaby, BC V5A~1S6, Canada}
	
	\author{Derek~Fujimoto}
	\altaffiliation[Current address: ]{TRIUMF, 4004 Wesbrook Mall, Vancouver, BC V6T~2A3, Canada}
	\affiliation{Department of Physics \& Astronomy, University of British Columbia, 6224 Agricultural Road, Vancouver, BC V6T~1Z1, Canada}
	\affiliation{Stewart Blusson Quantum Matter Institute, University of British Columbia, 2355 East Mall, Vancouver, BC V6T~1Z4, Canada}
	
	\author{Tobias~Junginger}
	\affiliation{TRIUMF, 4004 Wesbrook Mall, Vancouver, BC V6T~2A3, Canada}
	\affiliation{Department of Physics and Astronomy, University of Victoria, 3800 Finnerty Road, Victoria, BC V8P~5C2, Canada}
	
	\author{Victoria~L.~Karner}
	\altaffiliation[Current address: ]{TRIUMF, 4004 Wesbrook Mall, Vancouver, BC V6T~2A3, Canada}
	\affiliation{Stewart Blusson Quantum Matter Institute, University of British Columbia, 2355 East Mall, Vancouver, BC V6T~1Z4, Canada}
	\affiliation{Department of Chemistry, University of British Columbia, 2036 Main Mall, Vancouver, BC V6T~1Z1, Canada}

	\author{Spencer~Kiy}
	\affiliation{TRIUMF, 4004 Wesbrook Mall, Vancouver, BC V6T~2A3, Canada}
	
	\author{Ruohong~Li}
	\affiliation{TRIUMF, 4004 Wesbrook Mall, Vancouver, BC V6T~2A3, Canada}
	
	\author{Monika~Stachura}
	\affiliation{TRIUMF, 4004 Wesbrook Mall, Vancouver, BC V6T~2A3, Canada}
	\affiliation{Department of Chemistry, Simon Fraser University, 8888 University Drive, Burnaby, BC V5A~1S6, Canada}
	
	\author{John~O.~Ticknor}
	\affiliation{Stewart Blusson Quantum Matter Institute, University of British Columbia, 2355 East Mall, Vancouver, BC V6T~1Z4, Canada}
	\affiliation{Department of Chemistry, University of British Columbia, 2036 Main Mall, Vancouver, BC V6T~1Z1, Canada}
	
	\author{Robert~F.~Kiefl}
	\affiliation{Department of Physics \& Astronomy, University of British Columbia, 6224 Agricultural Road, Vancouver, BC V6T~1Z1, Canada}
	\affiliation{TRIUMF, 4004 Wesbrook Mall, Vancouver, BC V6T~2A3, Canada}
	\affiliation{Stewart Blusson Quantum Matter Institute, University of British Columbia, 2355 East Mall, Vancouver, BC V6T~1Z4, Canada}
	
	\author{W.~Andrew~MacFarlane}
	\affiliation{TRIUMF, 4004 Wesbrook Mall, Vancouver, BC V6T~2A3, Canada}
	\affiliation{Stewart Blusson Quantum Matter Institute, University of British Columbia, 2355 East Mall, Vancouver, BC V6T~1Z4, Canada}
	\affiliation{Department of Chemistry, University of British Columbia, 2036 Main Mall, Vancouver, BC V6T~1Z1, Canada}
	
	\author{Robert~E.~Laxdal}
	\email[E-mail: ]{lax@triumf.ca}
	\affiliation{TRIUMF, 4004 Wesbrook Mall, Vancouver, BC V6T~2A3, Canada}
	\affiliation{Department of Physics and Astronomy, University of Victoria, 3800 Finnerty Road, Victoria, BC V8P~5C2, Canada}
	
	\date{\today}
	
	\begin{abstract}
		A new high field spectrometer has been built to extend the capabilities of the \gls{bnmr} facility at TRIUMF.
		This new beamline extension allows \gls{bnmr} spectroscopy to be performed with fields up to \SI{200}{\milli\tesla}
		parallel to a sample's surface (perpendicular to the ion beam), allowing depth-resolved studies of local electromagnetic fields with spin polarized probes at a much higher applied magnetic field than previously available in this configuration.
		The primary motivation and application is to allow studies of \gls{srf} materials, close to the critical fields of \ch{Nb} metal, which is extensively used to fabricate \gls{srf} cavities. 
		The details of the design considerations and implementation of the \gls{uhv} system, ion optics, beam diagnostics  are presented here. 
		Commissioning of the beamline and spectrometer with radioactive ions are also reported here.
		Future capabilities and applications in other areas are also described.
	\end{abstract}
	
	\maketitle

	\section{
		Background and Motivation
		\label{sec:background}
	}

	Superconducting \ch{Nb} cavities are the enabling technology behind modern high-power and high-energy \glspl{linac}\cite{Padamsee2019,Padamsee2017}.
	Electromagnetic \gls{rf} fields are stored in \gls{rf} cavities and are used to efficiently accelerate charged particle bunches (or beams).
	To reduce \gls{linac} length, hence cost, the electric accelerating field can be increased, but with a proportional increase in the \gls{b-surf} parallel to the \gls{rf} cavity wall resulting in heat dissipation.
	Large oscillating $B$-fields can result in vortex generation and movement of those vortices generates heat as explained below.
	
	\gls{srf} \gls{linac} cavities (commonly made out of elemental \ch{Nb} metal) need to be operated in the superconducting state below the superconducting transition temperature $T_{c}$ (\SI{\sim 9.2}{\kelvin} for Nb), requiring extensive cryogenic infrastructure.
	\gls{srf} cavities can only be operated in the Meissner state when \gls{b-surf} is lower than the \gls{b-vp}, which for \ch{Nb} is on the order of \SI{200}{\milli\tesla}.
	In the Meissner state, magnetic fields are screened from the interior of the superconductor and can only penetrate on a \si{\nano\meter}-scale thin layer ($\sim$\SI{100}{\nano\meter} in \ch{Nb}) from the surface, defined by the \gls{lpd}.
	As magnetic fields are increased further beyond \gls{b-vp} in \ch{Nb} (a type-II superconductor), a transition from the Meissner state to the vortex state occurs where quantized magnetic fluxes (circulated by vortices of supercurrents) enter the interior of the superconductor.
	The rapid oscillation of these vortices in the \gls{rf} fields causes severe heat dissipation resulting in a sharp increase in the surface resistance\cite{Gurevich2017}. 
	
	\gls{rf} dissipation is contained within this surface layer and its response to the applied \gls{b-surf} has a significant impact on the overall performance of the \gls{linac}.
	Intensive research at \gls{srf} facilities worldwide has also demonstrated that the performance of \gls{srf} cavities is very sensitive to (and can be enhanced by) various surface modifications (e.g., vacuum heat treatment\cite{Ciovati2004, He2021, Ito2021, Lechner2021}, dilute impurity diffusion\cite{Grassellino2013,Grassellino2017}, and higher \gls{b-vp} superconductor thin film coatings\cite{Kubo2016,Gurevich2006,Gurevich2015,Posen2015}). 
	Theoretical studies postulate that a "dirty" layer on a "clean" substrate (or more complicated layered structures) can help shield the underlayer from the formation of dissipative vortices\cite{Checchin_GL_theory,Gurevich2017,Kubo2016}. 
	A deeper understanding of how the Meissner screening  deteriorates at \gls{b-surf} close to \gls{b-vp}, as well as how different surface modifications result in a reduced \gls{rf} dissipation and an enhanced \gls{b-vp} is required to improve cavity performance.
	
	\Gls{dc} magnetic fields applied both perpendicular and parallel to the sample surface have been used as analogues of cavity operating conditions to characterize \gls{srf} materials with respect to the field of first flux penetration\cite{Junginger2018}. Perpendicular fields result in highly non-uniform surface fields when the sample is in the Meissner state, with flux accentuated at sample edges. Parallel fields on coin shaped or ellipsoid samples provide near uniform surface fields up to the field of first flux penetration.  Measurements typically record the maximum local surface field (based on the applied field and the sample geometry) that can be reached before flux is detected in the bulk. Such measurements do not provide local details of the role of the surface layer(s) in the field of first flux entry. 
	Given the interest in layered structures and their precise role in enhancing the field of first flux entry, a diagnostic that provides the depth-resolved information of field attenuation through the London layer would provide new insight. 
	
	Useful tools to study \gls{srf} materials are techniques utilizing spin-polarized radioactive probes such as the \gls{mu} or radioactive \acrshort{li-8-neut} ions.
	In either case, the projectile is implanted into the bulk or near surface where the probe's spin reorients according to the static (i.e., time-average) and dynamic (i.e., stochastic) components of the local magnetic field.
	The $\beta$-decay emmissions of the probe are correlated with its spin orientation at the time of decay and the time evolution of the asymmetry of the $\beta$-decay provides information about the local magnetic field. 
	Early studies of \gls{srf} \ch{Nb} samples utilized the \gls{musr} facility at TRIUMF where magnetic fields were applied perpendicular to the face of the coin shaped samples of SRF material to characterize the field of first flux entry~\cite{Grassellino_muSR_TRIUMForig}. 
	Later, a spectrometer allowing the application of parallel fields was added\cite{Gheidi_SRF2015} and a number of measurements were performed both with ellipsoid and coin samples to characterize the role of various treatments on the field of first flux entry\cite{Junginger2018}. 
	The \gls{le-musr} facility at \gls{psi} has a much lower energy muon beam that allows depth-resolved studies in parallel fields~\cite{Salman2012,2008-Prokscha-NIMA-595-317,Morenzoni2000}. The low momentum muon projectile, however, is easily deflected by the applied parallel field on the sample and is limited to $\leq$\SI{30}{\milli\tesla}.    
	
	TRIUMF's implementation of the $\beta$-NMR technique~\cite{MacFarlane_ZPC2021,2015-MacFarlane-SSNMR-68-1} utilizes low-energy ($E\sim$ 20-30 \si{\kilo\electronvolt}) radioactive ions like \acrshort{li-8} that can be spin-polarized in-flight before delivery to a dedicated spectrometer.
	Like \gls{le-musr}, the \gls{bnmr} technique can be used to ``soft-land'' the ion at the near surface using a sample biased at \gls{hv}. The variable \gls{hv} bias is used to decelerate the ion to energies varying from hundreds of \si{\electronvolt} to the full ion energy which in turn allows depth-resolved studies of the local magnetic field. 
	The TRIUMF \gls{bnmr} facility\cite{2014-Morris-HI-225-173,2018-Kreitzman-JPSCP-21-011056} is capable of studying samples in high perpendicular field (up to \SI{9}{\tesla}) or (prior to this work) in parallel fields up to a maximum $\sim$\SI{24}{\milli\tesla}. 
	Compared to low energy muons, the transverse deflection of the \gls{rib} can be more easily compensated and therefore an upgrade of the TRIUMF parallel field capability was proposed and realized. 
	The new facility provides ion deceleration (for depth-resolved studies) at high parallel fields up to \SI{200}{\milli\tesla} --- a capability unique in the world.
	This paper describes the new facility, including the design considerations, the installed equipment, and the commissioning with both stable and radioactive beams.

	\section{
		\glsentryshort{bnmr} Technique and the Existing Facility at TRIUMF 
		\label{sec:existing-facility}
	}
	\gls{bnmr} is a  type of \gls{nmr} technique which utilizes implanted spin-polarized radioactive nuclei as probes.
	The main difference compared with conventional \gls{nmr} is the method of detection. 
	The local field can be monitored using the anisotropic emission of the $\beta$-particles (typically \si{\mega\electronvolt} energy), which is highly correlated with the direction of the nuclear spin at the time of the decay. 
	
	High energy $\beta$s are detected in two opposing simple detector telescopes (with pairs of scintillators in coincidence), oriented at 0 and $\pi$ radians relative to the direction of the initial spin polarization.
		Due to the parity violation of the $\beta$-decay, the $\beta$ counts per unit time $C_0$ and $C_\pi$ differ by a factor proportional to the angular emission probability $W(\theta)$, i.e.,
		\begin{align}
			C_\theta (t) &= \frac{1}{2}\frac{N(t)}{\tau}W(\theta) = \frac{1}{2} \frac{N(t)}{\tau}
			\begin{cases}
				\{1+A_0 P(t)\} &\text{for $\theta = 0$,}\\
				\{1-A_0 P(t)\} &\text{for $\theta = \pi$,}
			\end{cases}\label{eq:L_R_counter}
		\end{align}
		where the total number of nuclei (with mean lifetime $\tau$) present at time $t$ after the beam has turned on and implanted at a constant rate $R_0$, is defined as:
		\begin{equation}
			N(t) = \int_{0}^{t}N(t,t')dt' = R_0\int_{0}^{t} \exp[-(t-t')/\tau]dt'.\label{eq:N_t_count}
		\end{equation} 
		For the remaining terms in eq.~\ref{eq:L_R_counter}, $A_0$ is the proportionality constant which depends on both the properties (solid angle) of the detectors, ($\beta$ energy-dependent) detection efficiency, and the intrinsic asymmetry $A$ ($A_0\sim 0.1$ for \LiEight)\cite{MacFarlane_ZPC2021}. $P(t)$ in eq.~\ref{eq:L_R_counter} is the degree of spin-polarization.
		The count rates are combined to generate the experimental asymmetry:
		\begin{equation}
			A(t) \equiv \frac{C_0(t) - C_\pi(t)}{C_0(t) + C_\pi(t)} = A_0 P(t)\label{eq:asymmetry_count_rate},
		\end{equation}
		which is directly proportional to $P(t)$.
		The time evolution of the (longitudinal) spin-polarization can therefore be measured as the probe spin interacts with the local magnetic field inside the sample. 
		This method of detection allows about ten orders of magnitude higher sensitivity (per nuclear spin) than conventional \gls{nmr}, allowing for the study of situations that are either extremely difficult or impossible with conventional approaches (e.g., thin films, ultra-dilute impurities, etc.)~\cite{2014-Morris-HI-225-173,2015-MacFarlane-SSNMR-68-1,MacFarlane_ZPC2021}.  
		
	The implantation energy of \gls{bnmr} probes can be varied (e.g., from $\sim$\SI{0.5}{\kilo\electronvolt} to $\sim$\SI{30}{\kilo\electronvolt}) to study the surface of materials on the \si{\nano\meter}-scale and in a depth-resolved manner. 
	Similar to \gls{musr}, the \gls{bnmr} technique does not rely on the presence of a suitable \gls{nmr} nucleus in the material and can therefore be applied to any material at any applied magnetic field (including zero field). 
	In contrast to \gls{le-musr}, where the \gls{t-lifetime} is \SI{2.2}{\micro\second}, typical \gls{bnmr} probes have a much longer \gls{t-lifetime} (e.g., $\tau=$ \SI{1.21}{\second} for \LiEight~\cite{Flechard2010}). It is therefore more sensitive to dynamics on a much longer timescale, making the two techniques complementary (rather than competitive). 
	Furthermore, \gls{bnmr} ions at TRIUMF need to be spin-polarized in-flight via optical pumping, whereas surface \glspl{mu} produced from two-body pion decay are nearly 100\% spin-polarized. 
	
	The $\beta$-NMR technique as implemented at TRIUMF requires a dedicated facility to produce high intensity \glspl{rib} with a high degree of spin polarization.
	In the case of TRIUMF, the high intensity \glspl{rib} are produced at the \gls{isac} facility, which uses a \SI{500}{\mega\electronvolt} cyclotron as a driver to bombard a solid (radio)isotope production target with proton beams~\cite{Dilling_Hyperfine_ISAC}. 
	The resulting short-lived nuclides are then ionized, extracted, and separated on-line (i.e., using the so-called ISOL method) to produce isotopically pure \glspl{rib} before being sent downstream for various experiments. 
	\LiEightPlus\, beams with an intensity of \SI{\sim e7}{\per\second} are routinely available.

	Prior to being delivered to the \gls{bnmr} end-station, the \glspl{rib} at \gls{isac} are spin polarized in-flight using dedicated facility infrastructure (allowing for routine operation) via collinear optical pumping with a circularly polarized resonant laser~\cite{2014-Levy-HI-225-165}.
	This step produces \SI{\sim 70}{\percent} nuclear spin polarization~\cite{MacFarlane2014_InitSpinPol}. 
	The TRIUMF \gls{bnmr} facility has two beamlines: the \gls{bnmr} leg, which allows for measurements at high perpendicular fields up to \SI{9}{\tesla}, and the \gls{bnqr} leg, which allows samples to be studied in parallel fields up to $\sim$\SI{24}{\milli\tesla} (prior to this work). 
	The NQR technique is a zero field analogue of NMR that depends on the quadrupole interaction of nuclear spins $>$1/2 in solids.
	The spin-polarized \gls{rib} can be sent alternately to either of the two beamlines thanks to an electrostatic ``kicker'' that is used to ``pulse'' the beam.
	In this way two experiments can run simultaneously.
	
\begin{figure*}[hbt!]
		\centering
		\includegraphics[width=0.9\textwidth]{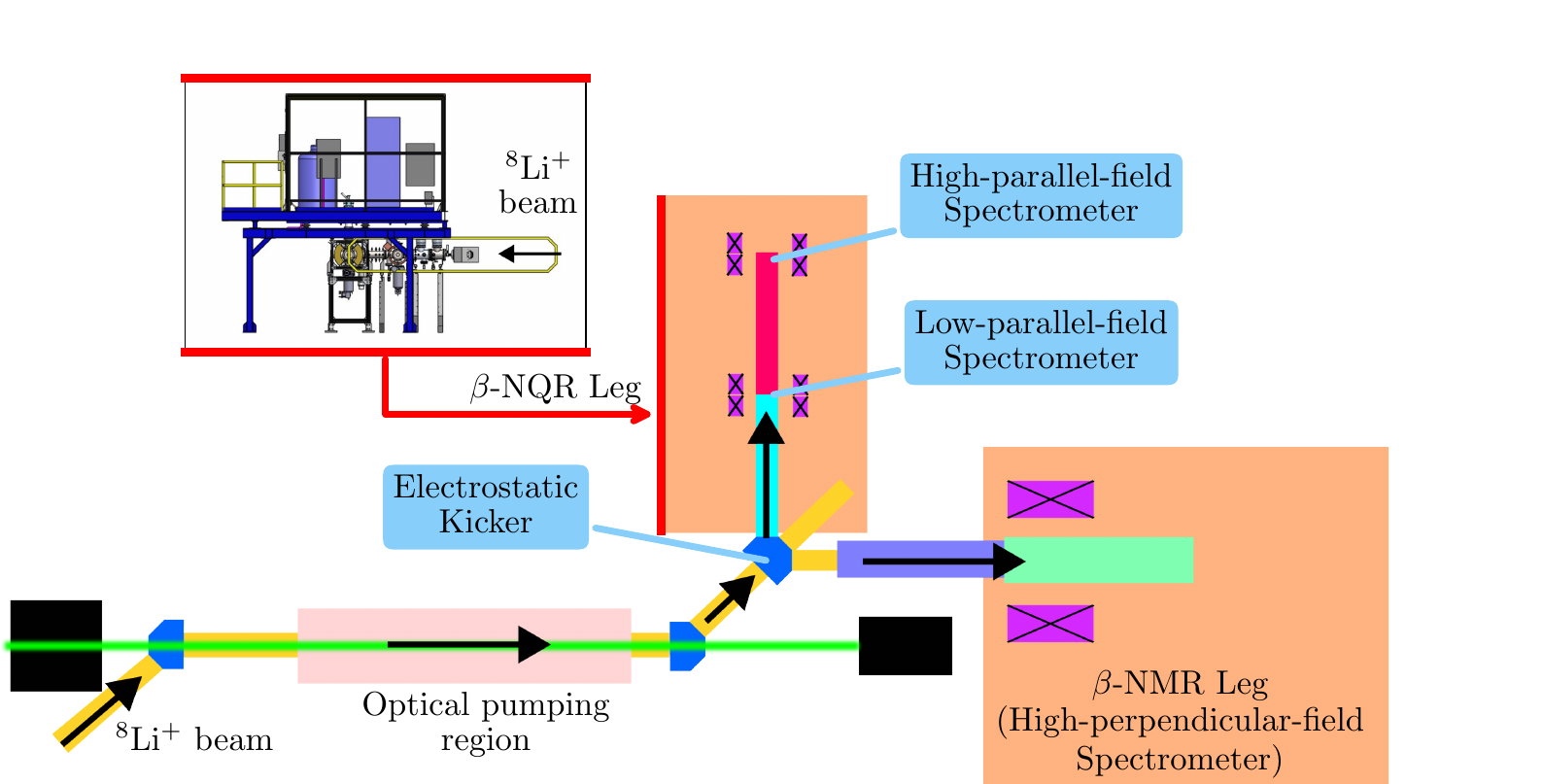}
		\caption{
			\label{fig:twospect}
			Beamline layout of the upgraded $\beta$-NMR facility. Typical beam energy extracted from the source for \gls{li-8} is $\sim$20-30 \si{\kilo\electronvolt}. The \gls{rib} is spin-polarized in-flight (first neutralized with an alkali vapour cell, and later re-ionized by a \ch{He} vapour cell) using dedicated optical pumping infrastructure, allowing for routine operation. The fast electrostatic kicker allows for semi-simultaneous operation of two spectrometers (i.e., two experiments running at the same time off of a single \gls{rib}). At the \gls{bnmr} leg, a spectrometer equipped with \gls{uhv} cold-finger cryostat and superconducting solenoid allows measurements at high magnetic fields (0.5-9 \si{\tesla}) perpendicular to the sample surface. The upgraded \gls{bnqr} leg provides a new spectrometer for measurements up to \SI{200}{\milli\tesla} parallel to the sample surface, just downstream of the existing low-parallel-field (0-24 \si{\milli\tesla}) one. The \gls{bnqr} spectrometer is also equipped with a \gls{uhv} cold-finger cryostat.  The inset shows the side view of the upgraded \gls{bnqr} beamline and the \gls{hv} platform above the beamline.
		}
	\end{figure*}

	At the end of each leg is a spectrometer assembly for sample measurements, containing the detectors, cold finger cryostat (for low temperature measurements), magnets, RF coil, \gls{hv} deceleration (\gls{hv} cage and deceleration electrodes), and \gls{ccd} imaging system. 
	In addition, in order to achieve depth-profiling the sample ladder and cryostat are raised to the bias of the \gls{hv} platform, which can be varied in order to decelerate the ion beam as it approaches the sample. 
	The bias capability spans from 0-35 \si{\kilo\volt} via a high stability (FuG Elektronik GmbH model HCL) \gls{dc} power supply.
	 A high voltage interlocked cage surrounds the biased equipment on a platform above the beamline (see the inset of fig.~\ref{fig:twospect}).
	
	The \gls{bnmr} beamline with high (perpendicular-to-sample-surface) field spectrometer uses a superconducting solenoid to generate fields up to \SI{9}{\tesla}. 
	The detectors are oriented forward and backward relative to the sample, and the emitted $\beta$-particles penetrate through the sample to reach the front detector located downstream of the sample.
	The back detector, located outside the \gls{hv} platform, contains a small aperture to allow passage of the \gls{rib}. 
	On the \gls{bnqr} beamline, the (parallel-field) spectrometer uses a normal conducting Helmholtz coil.
	The detectors are oriented to the left and right side of the sample, and up to four samples can be loaded simultaneously into the \acrfull{uhv} chamber via a sample ladder load-lock system.
	This paper reports the upgrade of the \gls{bnqr} leg with the addition of a 1m long extension of the beamline to include a parallel field spectrometer capable of up to \SI{200}{\milli\tesla} for unique depth-resolved studies of SRF and other materials.

	\section{
		High-Parallel-Field Upgrade
		\label{sec:technical-details}
	}
	
	The high-parallel-field upgrade on the \gls{bnqr} leg was completed in two phases. The scope of phase I was to modify and improve the existing low-parallel-field section. In phase II, an additional $\sim$\SI{1.02}{\meter} of new beamline was installed. Phase I and phase II were completed in June 2019 and in July 2021, respectively.

	\subsection{Ion optics}
\begin{figure*}[hbt!]
	\centering
	\includegraphics[width=0.9\textwidth]{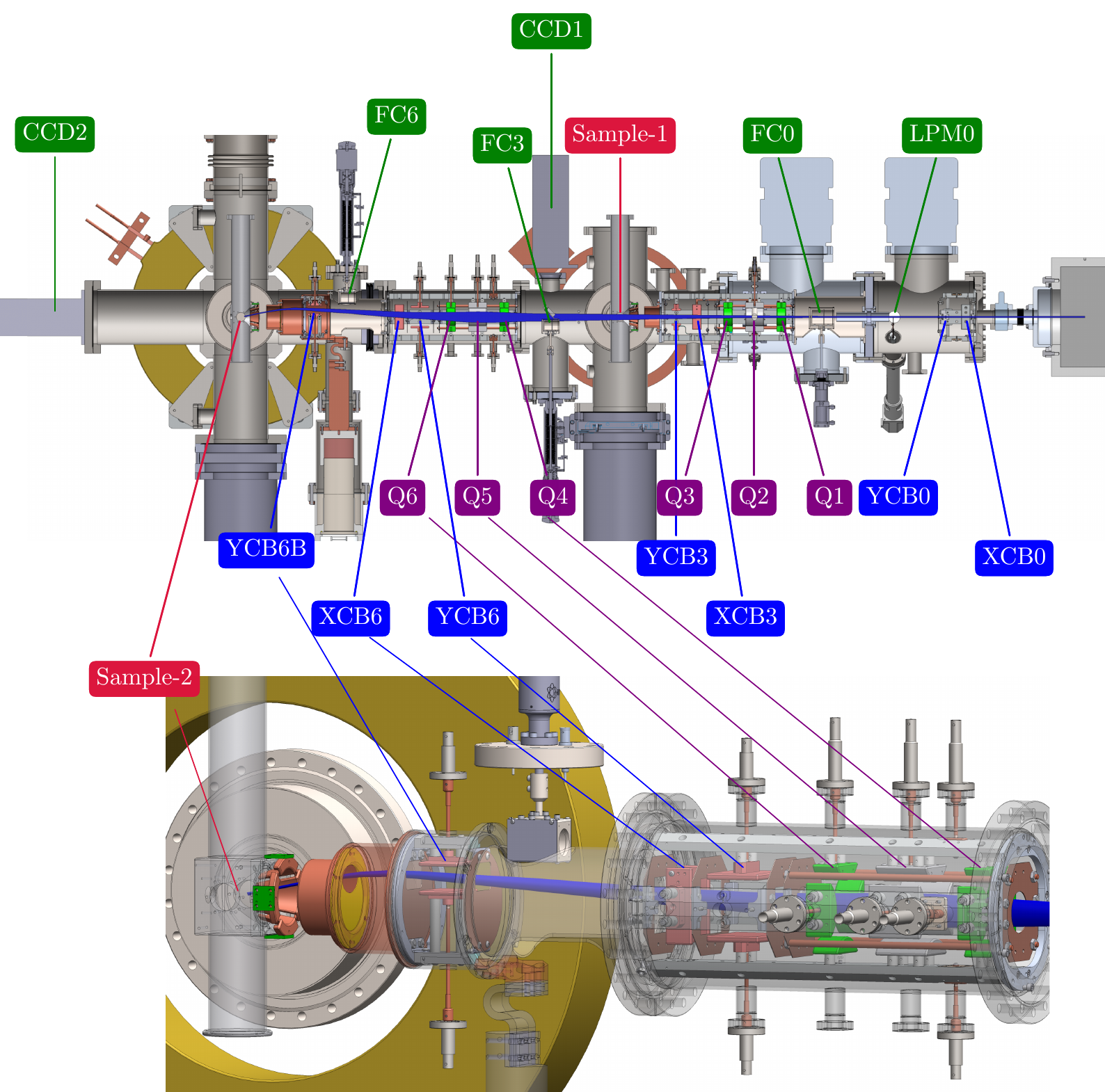}
	\caption{
		\label{fig:diagnostics}
		Details of the upgraded \gls{bnqr} leg (see also fig.~\ref{fig:twospect}). Top: The final straight section of the beamline (XCB and YCB refer to horizontal and vertical steerers, FC to Faraday cups, Q to quadrupoles, LPM to Linear Profile Monitor). 
		Bottom: Enlarged view of the new beam line extension model, showing the electrostatic quadrupoles, steerers, ``skimmer'' plates (grounded electrodes with a defined aperture to limit the effective length of optical element), Faraday cup, thermal radiation baffle and final quadrant decelerating electrode. A representative beam trajectory (blue envelope), obtained using a particle tracking code (General Particle Tracer\cite{GPT}) for a \SI{200}{\milli\tesla} applied field is illustrated in the model.
	}
\end{figure*}

	There are three main elements of ion optics used along the \gls{bnqr} beamline: electrostatic steerers, electrostatic quadrupoles, as well as a magnetic Helmholtz coil.
	The steerers compensate for the vertical deflection of the ions as they pass through the fringe of the applied magnetic field and deflect the beam onto the sample; the quadrupoles are used to control the beam shape during transport and to focus the ions onto the sample; and the Helmholtz coil provides the magnetic field on the sample parallel to the surface.
	An additional four-sector electrode close to the sample (within $\sim$\SI{50}{\milli\meter}) acts as a deceleration electrode, and allows independent horizontal and vertical steering which is used to compensate for the beam deflection and to re-center the beam on the sample.  
	The layout of the ion optics and Helmholtz coil in the \gls{bnqr} leg are shown in fig.~\ref{fig:diagnostics}. 
	There are 3 horizontal steerers (XCB) and 4 vertical steerers (YCB) and two sets of electrostatic quadrupole triplets.
	
Fig.~\ref{fig:diagnostics} also shows a model of the new beam line extension complete with a model beam trajectory for the case where the Helmholtz coil is producing \SI{200}{\milli\tesla} at the sample.
The magnetic fringe field diverts the beam $\sim$\SI{2.4}{\centi\meter} off-axis through a set of ``skimmers'', a vertical steerer, and a thermal radiation baffle. The vertical steerers adjust the position of the beam on the sample, while the radiation shield is used to reduce the heat load on the cryogenically cooled sample. 
Even though the design was focused on an applied field of 200 mT, the two vertical deflectors can be operated independently to achieve the desired beam path at lower fields (all the way down to zero-field).

\begin{figure*}[htb!]
	\centering
	\includegraphics[width=0.9\textwidth]{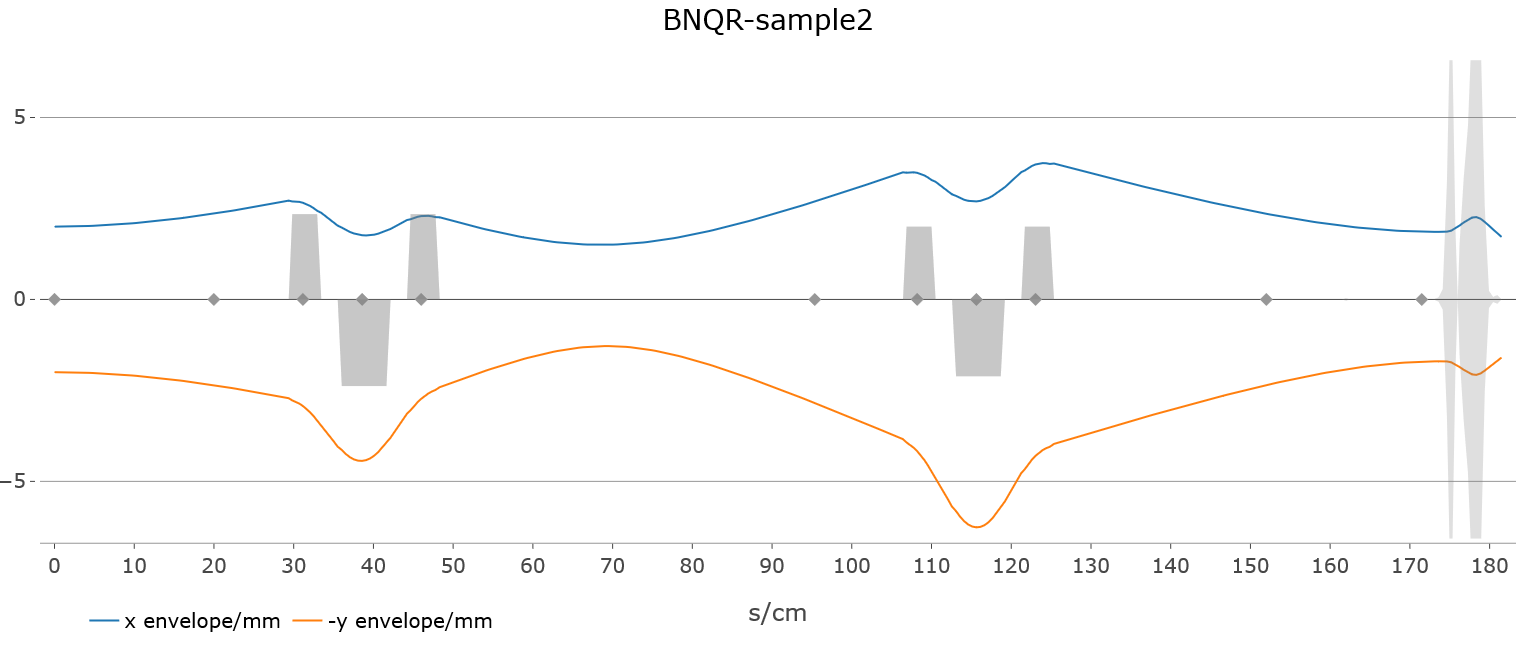}
	\caption{
		\label{fig:Envelope}
		Plot of the beamline envelope (beam going from left to right) obtained from the \emph{Envelope}
		calculator based on the TRANSOPTR code\cite{Heighway1981,TRANSOPTR_Manual}, used to tune the beam dimensionns under flexible operating conditions via two quadrupole triplets (i.e., for various beamspot sizes, beam energies, \gls{hv} platform biases, etc). The high-parallel-field and low-parallel-field spectrometer samples are located at \SI{181.5}{\centi\meter} and \SI{76}{\centi\meter} away from the \gls{lpm}0 (at \SI{0}{\centi\meter}), respectively. The amplitude of the shaded areas located around 39 \si{\centi\meter}, 115 \si{\centi\meter}, and 180 \si{\centi\meter} indicate the focal strength of the first quadrupole triplet, second quadrupole triplet, and \gls{hv} bias, respectively.
	}
\end{figure*}

Beam focusing is achieved using two sets of quadrupole triplets to obtain an axially symmetric beam on sample. 
Previously, an Einzel lens was used to focus the beam on the low-parallel-field (sample 1) spectrometer.
With the phase I upgrade, the first triplet replaced this Einzel lens and is now used to either obtain an optimum focus on the low-parallel-field spectrometer  or can be used in tandem with the second triplet to provide a variable symmetric spot size at the second sample position (sample 2).
The optimal setpoints for both triplets can be obtained using the TRANSOPTR code~\cite{Heighway1981,TRANSOPTR_Manual} via a web application called \emph{Envelope} (see fig.~\ref{fig:Envelope}), which is hosted on a local server at TRIUMF. 

Several \gls{hv} power supply modules providing bias to the electrostatic ion optics are installed in two separate crates. 
Both crates and all the plug-in modules were purchased from \acrlong{wiener}, with the first crate (8U or 8 standard rack unit, corresponding to 14" in height) installed in a 19" rack at the beamline level, and the second smaller crate (3U or 5.25" in height) installed in another 19" rack inside the \gls{hv} platform.
The larger crate on the beamline level houses five modules for the \gls{bnqr} leg ion optics requiring \gls{hv} bias relative to ground, and the smaller crate housing two modules is used exclusively for the four-sector electrode biased relative to the \gls{hv} platform.

\begin{figure}
	\centering
	\includegraphics[width=\columnwidth]{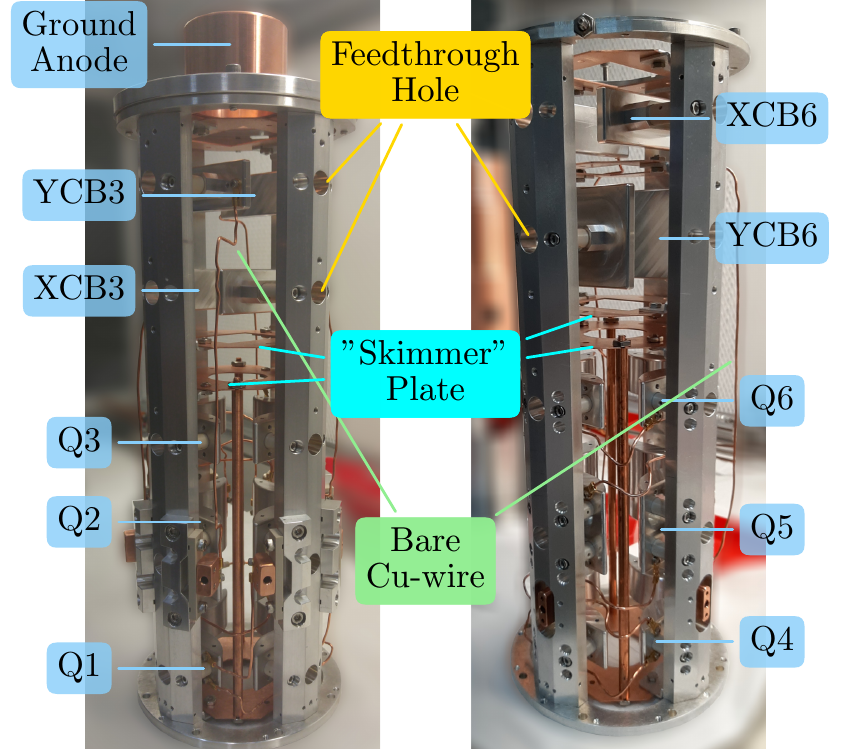}
	\caption{Ion optics ``boxes'' used in the upgraded \gls{bnqr} beamline. Feedthrough holes allow \ch{BeCu} extension posts connecting the \gls{uhv} feedthroughs to the \gls{hv} electrodes (see also details in fig.~\ref{fig:diagnostics}). The ``skimmer'' plates are grounded electrodes with apertures used to limit the effective length of the ion optics. Also shown are the stripped bare \ch{Cu} wires used to connect electrodes in parallel to the common \gls{uhv} feedthroughs. 
		Left: Ion optics box upstream of the low-parallel-field sample position, containing the first quadrupole triplet and steerer asssemblies. 
		Right: Ion optics box for the high-parallel-field section, containing the second quadrupole triplet and steerer assemblies. Note that the beam enters from the bottom side of the image.
		\label{fig:ion-optics-assembly}
	}
\end{figure}

For the larger crate, one bipolar module uses 10 \gls{hv} channels to provide up to $\pm$\SI{500}{\volt}  for the three horizontal steerers and the first two vertical steerers. Three additional unipolar modules provide six positive and six negative channels with biases up to \SI{6}{\kilo\volt} for the two quadrupole triplets.
All quadrupole electrodes with the same polarity are wired in parallel with \gls{uhv} compatible stripped, bare \ch{Cu} wires, as shown in fig.~\ref{fig:ion-optics-assembly}.
Not shown are the \gls{hv} feedthroughs installed in the external vacuum chamber with center conductors that connect to the internal \gls{hv} wires and electrodes (see also details in fig.~\ref{fig:diagnostics}).
One unipolar module (with 8 positive and 8 negative channels) uses 2 pairs of \gls{hv} positive and negative channels to provide up to \SI{3}{\kilo\volt} for each electrode of the last two vertical steerers, which is used to deflect the beam through the thermal radiation baffles.
Polarity reversal of this module is achieved with a switch. All beamline devices are remotely operated via an \gls{epics}~\cite{Dalesio1994,EPICS} interface.

	\subsection{Beam Diagnostics and Beam Tuning}
	
	Beam diagnostics are available at discrete locations to aid with the beam tuning  via measurements of both the beam current and the beam profile (see fig.~\ref{fig:diagnostics}).
	The beam current can be measured using three \glspl{fc}. The \glspl{fc} are driven by stepper motor actuators to allow precise positioning with respect to the beam trajectory.
	All the \glspl{fc} are typically biased between 60-350 \si{\volt} with either negative or positive polarity, to either trap secondary electrons (i.e., for an absolute current reading) or to amplify the beam intensity during operation (e.g., when using a low-intensity \gls{rib}).
	The driven position of the first two \glspl{fc} are aligned with the beamline center while the third \gls{fc} is positioned off-axis (from the beamline center) to allow maximum collection when the beam path follows the reference trajectory.
	The reference values for the optimum motor positions were obtained by measuring the alignment of the \glspl{fc} using a theodolite via direct line-of-sight  for the case where the Helmholtz coil is producing 200 mT at the sample. during beamline installation.
	
	The beam profile at the sample location is obtained using a \gls{ccd} camera (Starlight Xpress) mounted on a viewport, by imaging the light emitted by a scintillator (typically sapphire\cite{Salman2014_sapphire}) on the sample ladder.
	Imaging of the beam at the sample location is virtually limited to \gls{rib} as it is primarily the decay products and not the kinetic energy of the beam that induces scintillation (the $\beta$s and $\alpha$s produced during radioactive decay are in the \si{\mega\electronvolt} range, \num{\sim e3} times more energetic than the beam itself). 
	The beam profile (and centroid position) can also be measured using horizontal and vertical slits  driven at \ang{45} in combination with a downstream \gls{fc}, the so-called \acrfull{lpm}. 
	The \gls{lpm} plate includes two slits (horizontal and vertical with respect to the beam axis) and three circular collimators of different diameters (8.0 \si{\milli\meter}, 5.6 \si{\milli\meter}, and 4.0 \si{\milli\meter}) on a copper block that allows beam size and position definition for reproducible tuning.
	The \gls{lpm} is driven by a stepper motor, with the center of each slit or collimator (and corresponding stepper motor position) determined during the installation using a theodilite.
	
	The tuning strategy and use of the diagnostics is as follows. The quadrupole triplets are set to theoretical values (from the \emph{Envelope} web application) for a given beam energy and desired spot size. The vertical steerers are set to the theoretical value for the chosen Helmholtz coil setting (i.e., applied field). The beam is first centered on axis and aligned using the LPM0 aperture and downstream \glspl{fc}. 
	A pilot beam using a stable isotope (e.g., \LiSevenPlus) is often used to establish the trajectory with the aid of the \SI{25}{\milli\meter} diameter \glspl{fc}. Fine tuning of beam spot size and position, particularly at reduced energies, is achieved with direct beam spot imaging using a scintillator (sapphire\cite{Salman2014_sapphire}) and a \gls{ccd} camera which views the sample ladder through a glass viewport from the downstream side.
	The sapphire produces a bright image under irradiation by a \gls{rib}, even at the typical low beam intensity for \glspl{rib} of $10^6$ ions per second.
	Another scintillator material (YAP:Ce, Proteus, Inc.) has also been used to image stable beam (e.g., during beam development using a non-radioactive ion source) but with a much duller image even at \SI{5}{\micro\ampere}. 
	
	\subsection{Ultra-high Vacuum System}
	Ultra-high vacuum ($\sim 10^{-10}$ Torr) is essential to extend the lifetime of the clean surface of any sample under investigation, particularly during measurements at cryogenic temperatures, to prevent condensation of residual gases. 
	The UHV throughout the beamline is established by differential pumping using a single backing-pump and two turbo-pumps, which can provide pressures down to \SI{\sim e-9}{Torr}.
	Somewhat lower pressures at the two sample locations are achieved using two separate cryopumps, mounted directly below the sample cryostat, providing base pressures down to \SI{\sim e-11}{Torr}.
	
	The new high-parallel-field spectrometer is built on a stainless steel six-way cross with \gls{cf} flanges on each arm. 
	The diameter of the tube along the beam axis is 6" \gls{od}, which is a compromise between reasonable pumping speed and a more compact Helmholtz magnet. 
	The top port consists of an 8" \gls{cf} flange that provides access for the cryostat to be lowered into the beamline.
	The side ports are 10" \gls{cf} flanges on 8" \gls{od} tube stubs, onto which re-entrant ports with welded thin (\SI{0.05}{\milli\meter}) stainless windows are bolted.
	This window allows a large diameter space for mounting scintillation detectors within \SI{6.35}{\centi\meter} (2.5") of the sample.
	
	The three existing UHV chambers on the low-parallel-field section have been modified to accommodate the new quadrupole triplet, the \gls{lpm}, as well as the new \gls{fc} and viewport for the \gls{ccd} camera.
	Two additional vacuum chambers are also fabricated in-house for the beamline extension, which includes one six-way cross for the spectrometer chamber and one beamtube for the upstream steerers and quadrupoles. 
	All UHV chambers are electropolished after \gls{tig} welding and machining.
	Subsequently, the chambers are leak-checked, and measured with a \gls{rga} after beamline assembly and pump-down. 
	
	\subsection{
		Sample Environment
		\label{sec:beam-dynamics}
	}
	\subsubsection*{Magnet Coils}

\begin{figure}
		\centering
		\includegraphics[width=\columnwidth]{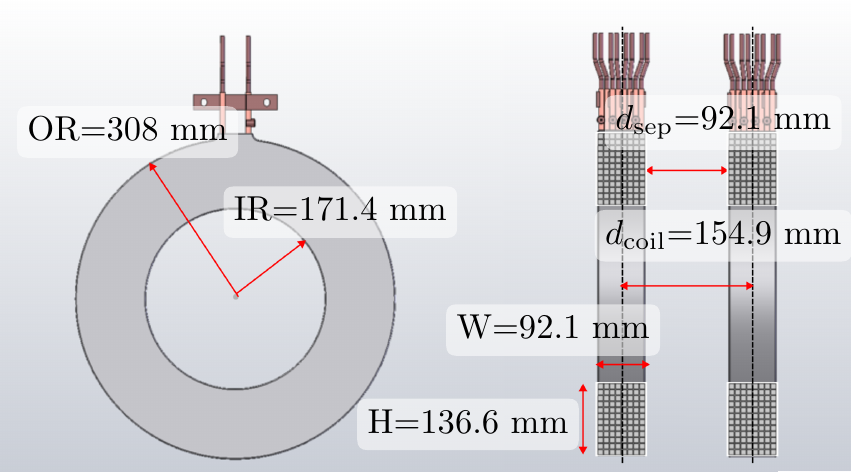}
		\caption{Dimensions of the \gls{bnqr} modified Helmholtz coils. Left: Inner radius (IR) and outer radius (OR) of a single coil in the high-parallel-field magnet installed at the sample 2 position. Also shown are the cooling channel tails at the top of the model. Right: The dimensions of the coil pair, modified from a Helmholtz configuration to fit the \gls{uhv} spectrometer chamber (with $d_\text{coil}$ as the central separation and $d_\text{sep}$ as the inner face separation between the coil pair). The width (W) and height (H) are the overall dimensions of the cross-section of the conductor turns.}
			\label{fig:Coil-dim}
		
	\end{figure}

	A copper conductor Helmholtz coil is used to provide a magnetic field of up to \SI{200}{\milli\tesla} at the sample position. 
	A field uniformity better than \SI{0.01}{\percent} is required  for \gls{bnmr} experiment across a typical beamspot size, \gls{fwhm} of $\sim$3-4 \si{\milli\meter}.
	The number of turns for each coil along the width and height is $8\times 12$, respectively, producing 56,080 \si{\ampere}-turns at a current of \SI{584}{\ampere}. The coil conductor is water-cooled with a 10 \si{\milli\meter} $\times$ 10 \si{\milli\meter} cross section from \gls{iacs}, consisting of \SI{100}{\percent} \ch{Cu}. Turn-to-turn insulation is provided by \SI{0.5}{\milli\meter} thickness double Dacron\textregistered\,glass, and \SI{1}{\milli\meter} thickness of additional fiberglass insulation. Each coil is composed of stacks of four pancakes encapsulated with vacuum impregnated clear epoxy resin. The resulting dimensions for a single coil are shown in fig.~\ref{fig:Coil-dim}, with a separation distance (between inner face of each coil) of \SI{154.9}{\milli\meter}. 
	The specified coils are sourced from Stangenes Industries Inc. (Palo Alto, CA).
	
\begin{figure}
	\centering
	\includegraphics[width=1.0\columnwidth]{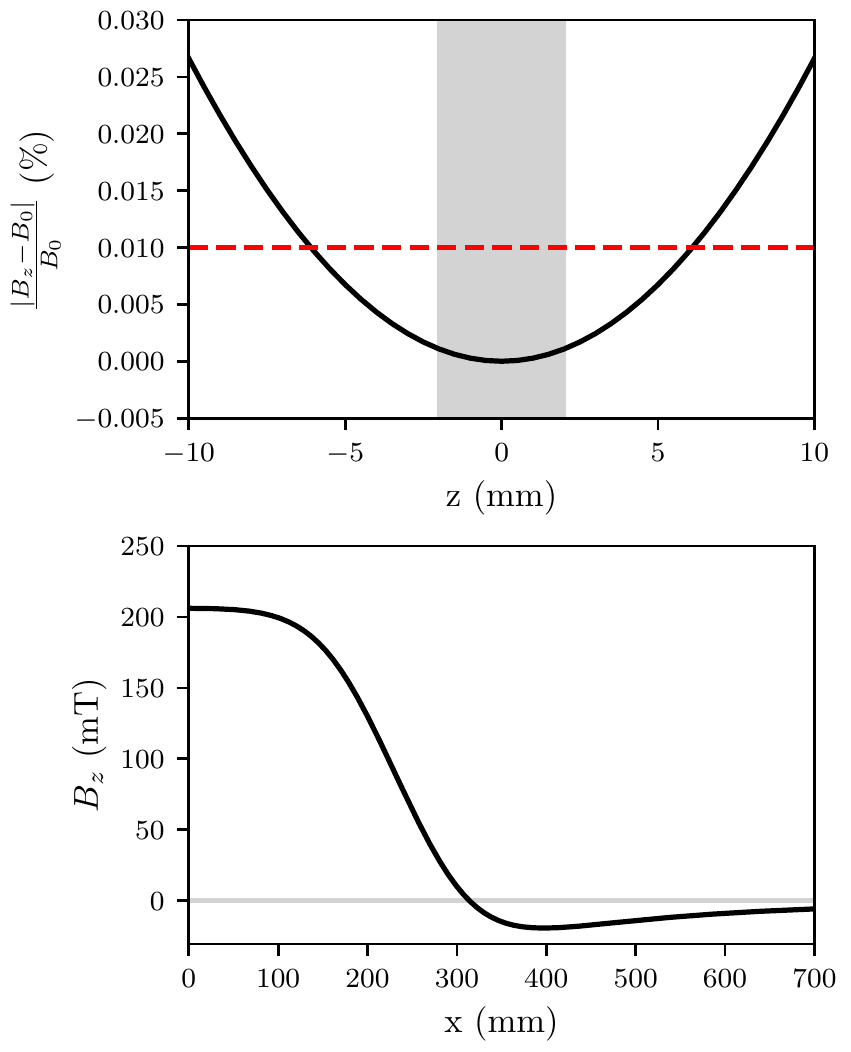}
	\caption{Top: Relative field variation along the Helmholtz coil axis (i.e., across the sample) obtained by summing the measured field profile of a single coil. The sample is centered at $z = \SI{0}{\milli\meter}$, where the uniformity of the applied field (\SI{206}{\milli\tesla} here) is maximal.
	The red dashed line indicates the required field uniformity across the typical \gls{fwhm} of the beamspot (grey shaded area). 
	Bottom: Calculated stray fields (transverse to the beam momentum and parallel to sample surface) along the beamline axis (i.e., the $x$-axis) from \gls{fem} simulation. The sample is located at the $x = \SI{0}{\milli\meter}$.
		\label{fig:Coil-fields}
	}
\end{figure}

\begin{figure}
	\centering
	\includegraphics[width=1.0\columnwidth]{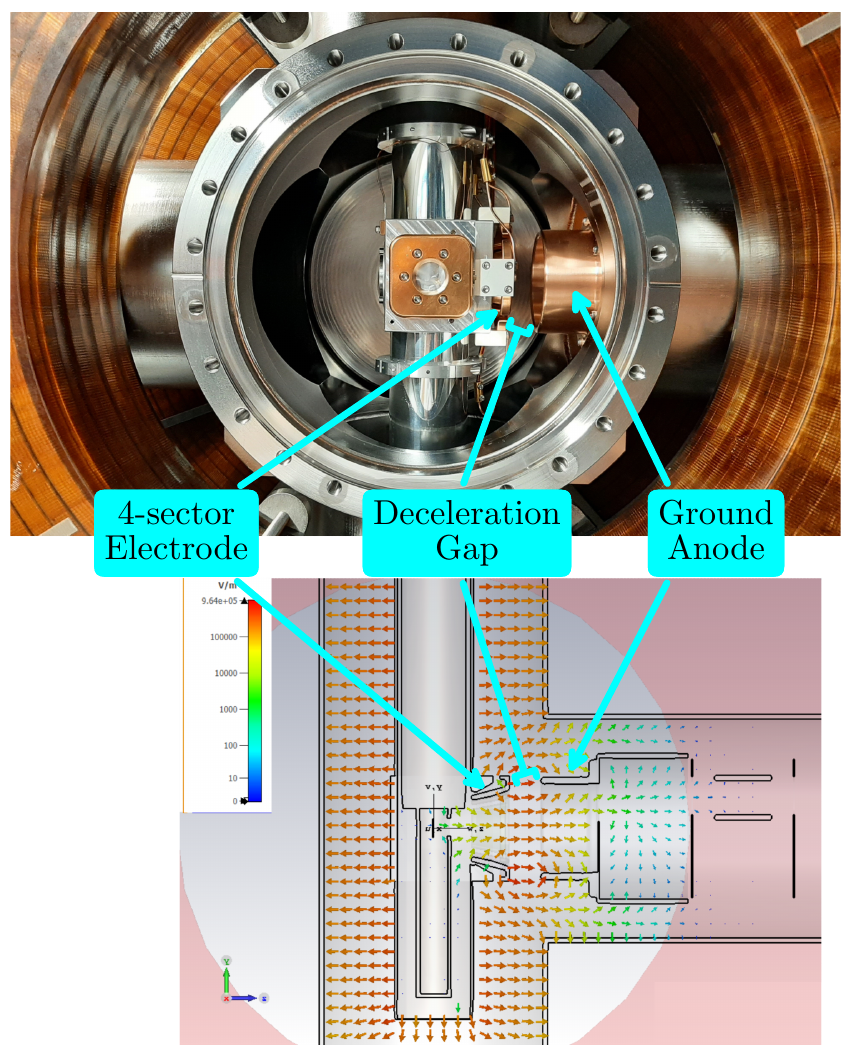}
	\caption{Top: View of the \gls{bnqr} spectrometer through one of the beamline viewports with the stainless steel detector window removed. Note that the beam enters from the right side of the image. Shown here are the cryostat, the four-sector electrode (mounted on the cryostat heat-shield) and the \SI{200}{\milli\tesla} Helmholtz coil. Bottom: Calculated electric fields (using CST Studio)\cite{CST} with \SI{10}{\kilo\volt} bias applied to the sample (and the cryostat). The beam is decelerated to the desired implantation energy within a small (\SI{\sim 21}{\milli\meter}) gap between the ground anode and the four-sector electrode.} 
	\label{fig:spectrometer-open}. 
\end{figure}

The magnetic fields along the coil pair axis (i.e., the parallel fields on the sample) and along the beamline axis were calculated using the \gls{fem} software OPERA-3d (TOSCA Solver for Static Current Flow problems)\cite{OPERA}. The simulated magnetic field homogeneity parallel to the sample surface (along the coil pair axis) and fringe fields along the beam axis are shown in fig.~\ref{fig:Coil-fields}. 
The power supply provides up to \SI{600}{\ampere} at \SI{50}{\volt} and is specified to have a stability of 0.01\% ripple  (\SI{60}{\milli\ampere} rms) with a digital resolution of 16-bit at \SI{200}{\milli\tesla}.
The power supply is sourced from Alpha Scientific Electronic (Stanton, CA) and installed with a water cooling system.
Remote control of the power supply is available through an \gls{epics} interface.~\cite{Dalesio1994,EPICS}
A photograph of the magnet installed around the cryostat is shown in fig.~\ref{fig:spectrometer-open}. The cryostat is biased at the platform voltage and the copper electrode to the right of center is the grounded anode defining the deceleration gap.

\subsubsection*{Cryostat Modification}
The existing cold-finger cryostat was modified to allow application of \gls{hv} bias for the four-sector electrode which is mounted on the cryostat radiation shields via SHAPAL\textsuperscript{TM} Hi-M soft spacers, the latter providing electrical isolation while maintaining high thermal conductivity. 
The cryostat and the HV power supply for the four-sector steerers are housed inside the existing electrically isolated \gls{hv} platform (via a ceramic vacuum break) to allow \gls{hv} bias for beam deceleration.
\Gls{uhv} compatible wires, such as Kapton insulated coaxial cables and \gls{ofhc} bare copper wires, and lead-free solder joints are used for the assembly.
Non-magnetic components such as connectors and fasteners are also used in the vicinity of the sample. 

	To accommodate for thermal contraction during low temperature measurements, the cryostat is supported on motorized bellows, which can be used to adjust the vertical alignment of the sample. In practice, the alignment is done by using the CCD camera at lower temperature to determine the current position and compare to a reference image with optimal alignment (usually determined at room temperature).
	The existing \gls{hv} platform is also modified to accommodate a new support structure for the cryostat.
	
\begin{figure*}[hbt!]
	\centering
	\includegraphics[width=0.85\textwidth]{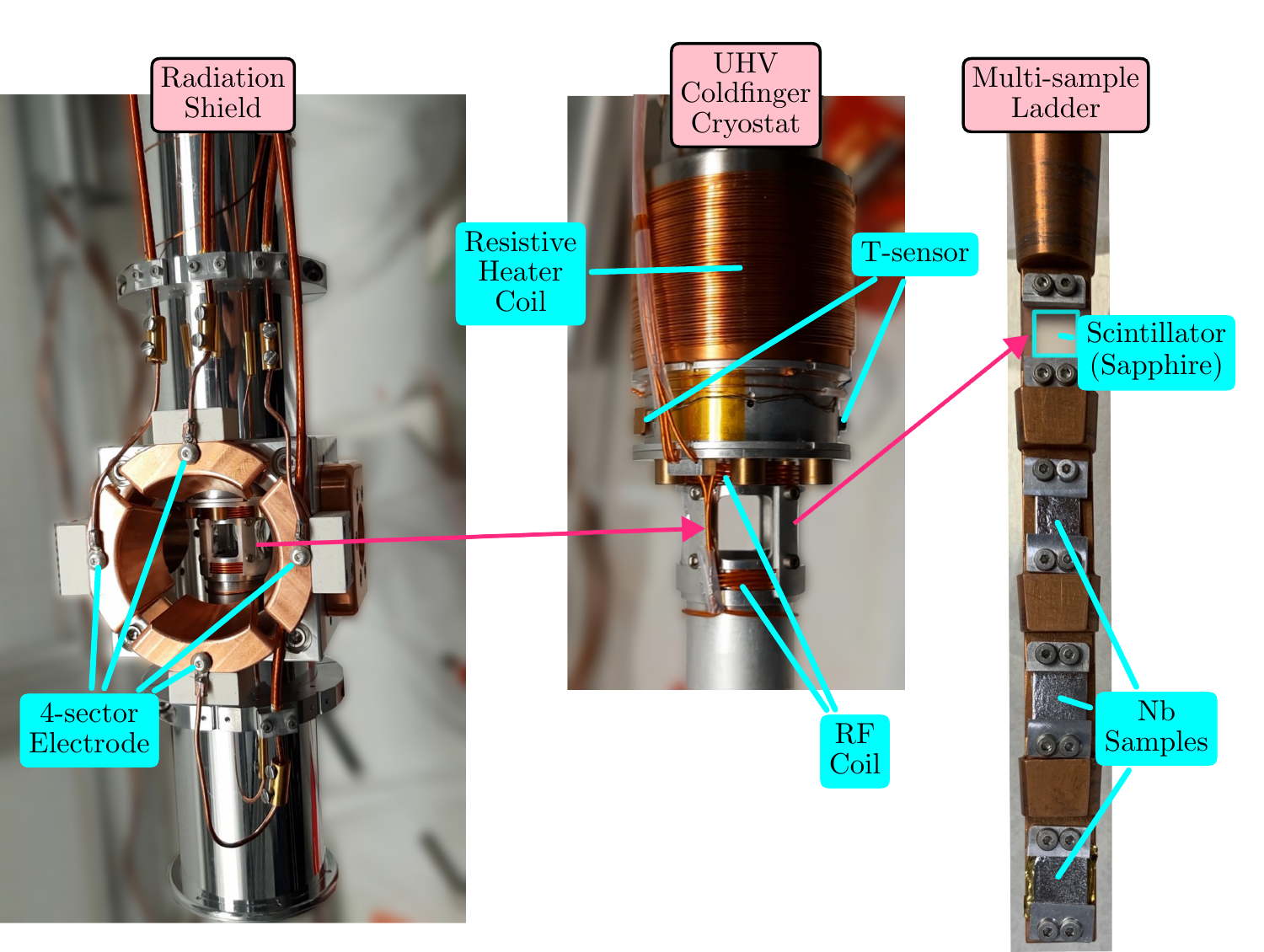}
	\caption{
		Left: Radiation shield of the $\beta$-NQR cryostat with the four-sector electrode shown facing towards the incoming beam. Middle: Sample environment inside of the cryostat (with the radiation shield removed). Right: The sample ladder which is used to mount up to four samples via a \gls{uhv} load-lock. The first position of the sample ladder in the photo is reserved for a sapphire scintillator, used for beam imaging using a CCD camera. The cyan-colored border indicates a 8 \si{\milli\meter} $\times$ 8 \si{\milli\meter} viewing window of the CCD camera through the back of the sample ladder.\label{fig:sample_config}.}
\end{figure*}

A continuous flow of \gls{LHe} can be provided via a portable \gls{LHe} dewar, which is installed onto the HV platform prior to the experiment.
Operating temperatures between \numrange[range-phrase=--]{4.5}{300}\si{\kelvin} are routinely achievable.
Two temperature sensors (\ch{GaAlAs} diodes from Lake Shore Cryotronics, Inc.) are mounted on the side of an \ch{Al} ring that holds the sample ladder.
An additional platinum sensor (\ch{Pt}-100 from Lake Shore Cryotronics, Inc.) is also installed on the heat shield.
The temperature is controlled by adjusting the  \gls{LHe} transfer line needle valve, the \ch{He} vapor mass flow to a roughing pump at the exhaust port of the cryostat, and the heater current for the cryostat (the latter is shown in fig.~\ref{fig:sample_config}).
Temperature stability is achieved with a Lake Shore model 336 controller driving a resistive heater on the cryostat cold block.
Operating at colder sample temperatures ($<$\SI{3}{\kelvin}) is foreseen in the near future with the addition of a cryo-head to be connected to the radiation shield. 
Measurements at even lower temperatures are planned in the future with the addition of a \ch{He}-3 cryostat. 

	\subsubsection*{Detector System}
\begin{figure}
		\centering
		\includegraphics[width=1.0\columnwidth]{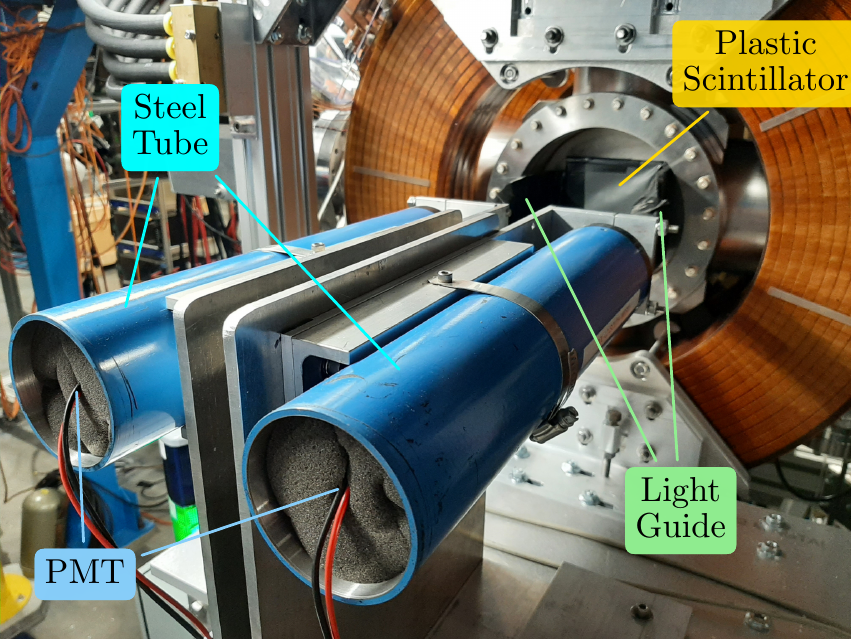}
		\caption{The detector assembly, consisting of the BC412 plastic scintillator, Lucite light-guide, and the PMTs (mounted in single steel tubes) on their mechanical support.
			\label{fig:spectrometer-closed}}
	\end{figure}

	Two scintillator telescope pairs (two on the left and two on the right of the cryostat position) are used to ``view'' the sample through thin (\SI{0.05}{\milli\meter}) stainless steel windows.
	The scintillator pairs are spaced sufficiently far apart to define a solid angle accepting $\beta$s originating at the sample, but rejecting most cosmic events and any $\beta$s from ions stopped in nearby parts. 
	For the high-parallel-field spectrometer, Hamamatsu H6153-01 fine mesh dynode type \glspl{pmt} are used, as they are more resistant to the relatively high magnetic fields used during the experiments. 
	The detector assembly, consisting of the scintillator (Saint-Gobain BC412 plastic scintillator with dimensions of 10 \si{\centi\meter} $\times$ 10 \si{\centi\meter} $\times$ 0.3 \si{\centi\meter}), Lucite light-guide, and the PMTs mounted in single steel tubes that provide limited magnetic shielding, are shown in fig.~\ref{fig:spectrometer-closed}.
	
	\section{
		Commissioning and first results
		\label{sec:comissioning}
	}
	As part of the instrument's commissioning, all devices, including their control interfaces and interlock logic, are tested without beam to establish that functional requirements are met. 
	\textit{In operando} tests using ion beams are organized into stable beam tests using light ions (e.g., \LiSevenPlus\, and \CTwelvePlus) at \SI{20}{\kilo\electronvolt} and \gls{rib} tests using \LiEightPlus\, at \SI{20}{\kilo\electronvolt}. 
	The stable beam tests are done to check that the optics (i.e., quadrupoles, steerers, etc.) are manipulating the beam as expected. The \gls{rib} is used to further check functionality, such as the final beam spot at the sample in order to confirm the optics and to check the detectors and \gls{daq} system. The Helmholtz coil field strength and direction are checked with a hand held probe (Gauss meter). A final confirmation is done by using polarized \LiEight\, beam and using the \gls{bnqr} signal to directly measure the magnetic field on the sample (as described below).

	There are two basic types of \gls{bnmr} measurements: resonance and relaxation. The beam commissioning with \gls{rib} employed both methods to check the full functionality of the equipment.
	A resonance measurement seeks to find the frequency (i.e., the resonance condition) that corresponds to an energy difference between a pair of the probe's magnetic sublevels.
		At TRIUMF, resonances are performed (predominantly) using continuous beam delivery and a transverse \gls{cw} RF field $B_{1}$ to manipulate the probe spins. 
		In the \gls{cw} approach, the RF field is stepped slowly through a range of frequencies close to the probe's Larmor frequency:
		\begin{equation}
			\nu_{0} = \frac{\omega_{0}}{2\pi} = \frac{\gamma}{2\pi}B_{0} ,\label{eq:resonant_freq}
		\end{equation}
		where $\gamma / (2 \pi) = \SI{6.30221 \pm 0.00007}{\mega\hertz/\tesla}$~\cite{IAEA_TableNuclearMom_gyro_Stone2019} is the \LiEight\, gyromagnetic ratio and $B_{0}$ is the applied magnetic field.
		On resonance, in the limit of sufficiently large $B_{1}$,  the populations of the sublevels involved in the transition(s) become equalized, resulting in a reduction in spin-polarization (or asymmetry).
	
	In a relaxation measurement, the temporal evolution of the probe's spin polarization (which is initially very far from thermal equilibrium) is monitored as it returns to thermal equilibrium. 
	The most commonly used type of such measurements is a so-called spin-lattice relaxation (SLR) measurement, wherein the applied field is parallel to the initial spin-polarization direction. 
		Spin-polarization of the probe is lost through an energy exchange with its surrounding environment (often called the ``lattice'' in NMR literature)\cite{SlichterBook}. 
		That is, spontaneous, stochastic fluctuations in the local field that are transverse to the probe spin direction serve to ``reorient'' it back to thermal equilibrium.
	Real time data for either mode can be conveniently displayed during the running experiment and analyzed using specialized \gls{bnmr} analysis software such as bfit\cite{bfit_arXiV,bfit_paper}. 

\begin{figure}
		\centering
		\includegraphics[width=1.0\columnwidth]{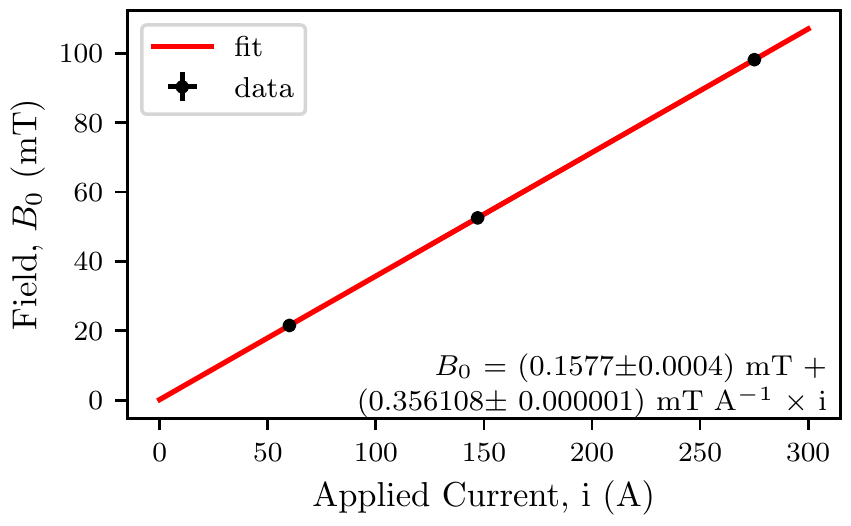}
		\caption{
			\label{fig:field-calibration} Calibration of the high-parallel-field coil via the \LiEight\,\gls{nmr} frequency in \ch{Au} foil at different currents applied to the Helmholtz coils. The solid red line denotes a linear fit to the data, whose expression appears in the inset.
		}
	\end{figure}

	The following calibration measurements were used to check the functionality of the detector, new Helmholtz coil, and the \gls{daq} system. 
	The magnetic field of the new coil on the sample location was precisely measured using the \acrshort{li-8-neut} resonance (see above description). 
	A 99.99\% pure \ch{Au} foil (Alfa Aesar) was chosen as the calibration sample due to its relatively narrow resonance line and slow spin-lattice relaxation rate down to very low magnetic fields.~\cite{MacFarlane_Gold,Parolin_Gold}
	The measured magnetic field on the \ch{Au} foil can be obtained from the frequency of the resonant peak, $\nu_{0}$ (eq.~\ref{eq:resonant_freq}), as a function of the current in the Helmholtz coil.
	The results are shown in fig.~\ref{fig:field-calibration}, demonstrating a linear relation between the applied current and the measured magnetic fields, as expected.
	For this calibration, stray fields originating from the superconducting solenoid at the \gls{bnmr} leg (operated at \SI{2.2}{\tesla} during calibration measurements) contribute to a small constant background field.
	
\begin{figure}
		\centering
		\includegraphics[width=0.7\columnwidth]{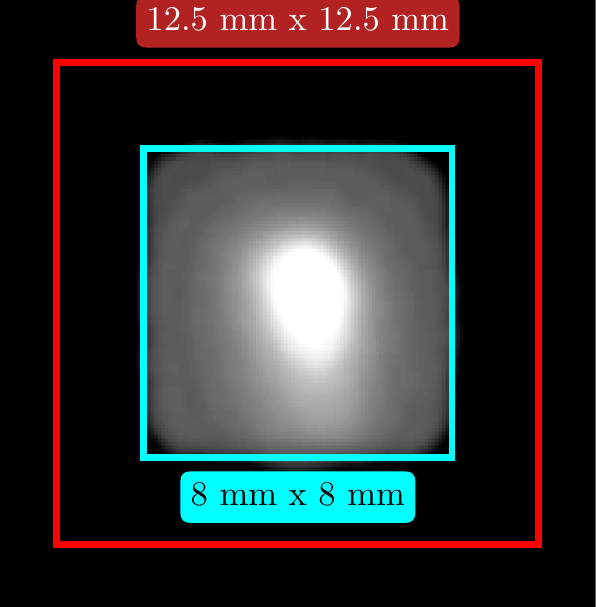}
		\caption{
			\label{fig:beamspot}
			Beamspot obtained at \SI{200}{\milli\tesla} on a sapphire scintillator. The image shown is looking upstream from the back side of the scintillator, viewed using the downstream CCD camera mounted on a viewport. The small (cyan) bounding box  corresponds to the dimension of the visible area of the sapphire from the aperture at back of the sample ladder (see fig.~\ref{fig:sample_config}), while the larger (red) box indicates the real size of the scintillator.
		}
	\end{figure}

	During commissioning with \gls{rib}, various tunes (ion optics values along the beam path) at \SI{20}{\kilo\electronvolt} \gls{li-8} are established  and their record is stored in a local database. 
	These tunes can be re-scaled for different beam energies, applied magnetic fields,  and \glspl{rib}, as well as re-loaded (i.e., as a starting condition) to speedup future beam delivery. 
	An example of a typical beamspot image is shown in fig.~\ref{fig:beamspot}, obtained at \SI{200}{\milli\tesla} using a 12.5 \si{\milli\meter} $\times$ 12.5 \si{\milli\meter} sapphire plate at one of the sample positions on the sample ladder.
	The average transverse beamspot dimensions (i.e., \gls{fwhm}) at the sample position for various applied magnetic fields are determined to be $\sim$3-4 \si{\milli\meter}.
	
\begin{figure}[hbt!]
		\centering
		\includegraphics[width=\columnwidth]{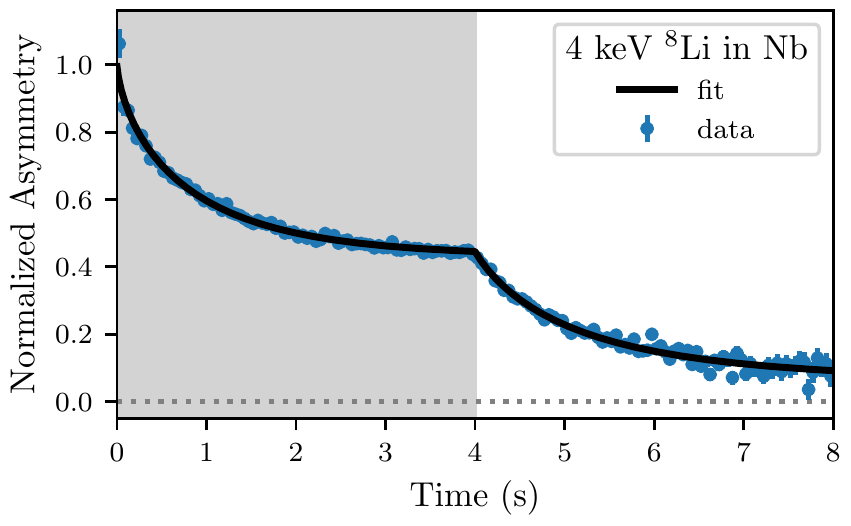}
		\caption{
			\label{fig:nb-slr}
			\LiEight\,\gls{slr} normalized asymmetry spectrum, i.e., $A(t=0)=1$, measured at \SI{4.3}{\kelvin} and \SI{200}{\milli\tesla} applied field on a \ch{Nb} sample using \SI{4}{\second} beam pulses of \SI{4}{\kilo\electronvolt} \LiEightPlus.  The shaded area indicates the measured asymmetry during beam pulse on. The solid line is the normalized fit to eq.~\ref{eq:SLR_fitfunc}.
		}
	\end{figure}

	Currently, \gls{slr} measurements on superconducting \ch{Nb} samples have been successfully performed at several magnetic fields up to \SI{200}{\milli\tesla}.
	The \gls{slr} rate, commonly denoted as $1/T_1$, characterizes the time-constant in depolarization of the \LiEight\, nuclear spins after they are stopped inside the sample. The value of $1/T_1$ is extracted by fitting the measured time-dependent asymmetry $A(t)$ with a phenomenological depolarization function, e.g., $p(t,t';1/T_1) = \exp[-(t-t')/T_1]$ for a single exponential depolarization function, which is convolved with the square or rectangular beam pulse:
		\begin{widetext}
			\begin{align}
				A(t) = A_0 
				\begin{cases} 
					\cfrac{R_0 \int_{0}^{t}\exp[-(t-t')/\tau] p(t,t';1/T_1) dt'}{N(t)}  &\text{for $t\leq\Delta$,}  \\
					\cfrac{R_0 \int_{0}^{\Delta}\exp[-(t-t')/\tau] p(t,t';1/T_1) dt'}{N(\Delta)}      & \text{for $t>\Delta$,}
				\end{cases}
				\label{eq:SLR_fitfunc}
			\end{align}
	\end{widetext}
where $\Delta$ is the beam pulse length (typically $\sim$1-4 \si{\second}), $\tau$ is the $^8$Li lifetime, and $N(t)$ is the total count rate as defined earlier in eq.~\ref{eq:N_t_count}. 
	Fig.~\ref{fig:nb-slr} shows a typical asymmetry spectrum and its fit to eq.~\ref{eq:SLR_fitfunc}, measured on a \gls{RRR}$\sim$300 \ch{Nb} sample (cut, etched, and annealed at \SI{1400}{\celsius}) at \SI{5}{\kelvin} using a \SI{4}{\kilo\electronvolt} \gls{li-8} beam. 
	Applications are available to display the raw counts, the calculated asymmetry, and a fit to various depolarization functions during the experiment\cite{bfit_arXiV,bfit_paper}.
	
\begin{figure}
		\centering
		\includegraphics[width=\columnwidth]{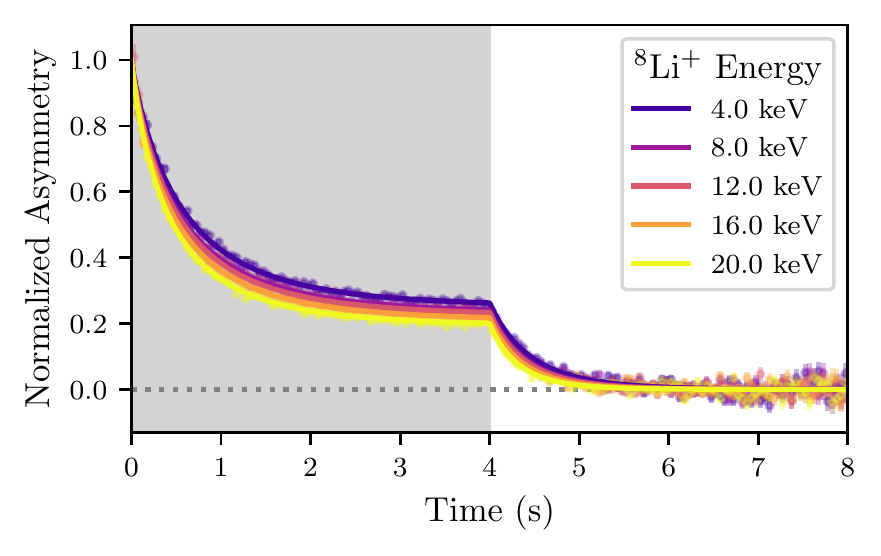}
		\caption{
			\label{fig:nb-scds}
			Normalized fits to the \gls{slr} asymmetry spectra, i.e., $A(t=0)=1$, of a \ch{Nb} sample (\gls{RRR}$\sim$300) at various implantation energies measured at \SI{4.5}{\kelvin} and at \SI{98}{\milli\tesla}. 
			The higher relaxation rates deeper below the surface reflects the reduced magnetic field due to Meissner screening. 
		}
	\end{figure}

	At applied magnetic fields below $\sim$\SI{1}{\tesla}, the dominant relaxation mechanism in \ch{Nb} is due to cross relaxation between the \LiEight\, and the 100\% abundant host \NbNinetyThree\, nuclear spins\cite{Parolin_Nb}, which gives a Lorentzian dependence of $1/T_1$ on the local magnetic field according to:
	\begin{equation}  
		\frac{1}{T_1} \approx \frac{ ( \gamma B_d )^2 \cdot ( 1/\tau_c ) }{ (1 / \tau_c)^2 + [ \gamma B_{\text{loc}} ]^2 } ,\label{eq:SLR_Rate_Redfield}
	\end{equation}
	where $\gamma$ is the aforementioned \LiEight\, gyromagnetic ratio, $B_\text{loc}$ is the measured local magnetic field, $B_d$ is the magnitude of the fluctuating dipolar field (due to the host \NbNinetyThree\, nuclear spins), and $\tau_c$ is the correlation time for the fluctuation in $B_d$. 
	The variation of magnetic fields in the superconducting \ch{Nb} due to Meissner screening can therefore be measured from the asymmetry spectra at different implantation energies (which corresponds to different implantation depths).
	Experimental results on the same sample shown in fig.~\ref{fig:nb-slr} for five different implantation energies are shown in fig.~\ref{fig:nb-scds}.
	Here, a \ch{Nb} sample is probed by \LiEight\, at five different depth distributions corresponding to ion energies of 4, 8, 12, 16 and 20 \si{\kilo\electronvolt} and average depths of 12, 22.5, 33, 44, 55 \si{\nano\meter}, respectively. 
	The average depths are computed using the \gls{srim} package~\cite{2010SRIM}.
	The data set shows the expected signature of Meissner screening in superconducting \ch{Nb}\cite{Hossain2009} as the depolarization rate increases as the ion is implanted deeper into the surface.
	Obtaining the accurate values of the local magnetic field requires a detailed analysis, taking into account the implantation distribution of the \gls{li-8} ions at various energies, which averages over fields at different depths.
	The details of this analysis will be reported in a separate publication.
	
\section{
	Scientific Applications \& Future Capabilities
	\label{sec:scientific-applications}
}

Just like conventional NMR using stable nuclei,
the breadth of applications of $\beta$-NMR is enormous.~\cite{2015-MacFarlane-SSNMR-68-1,MacFarlane_ZPC2021,2018-Kreitzman-JPSCP-21-011056,2014-Morris-HI-225-173} 
The ability to operate at intermediate magnetic fields
(10s to 100s of \si{\milli\tesla})
opens new opportunities for scientific applications.
In addition to the \gls{srf} material investigations, some of the experiments enabled in this field regime are: studies of thermally activated dynamics (i.e., moving the Bloembergen-Purcell-Pound peak\cite{BPP_orig} via adjusting $B_{0}$)
~\cite{2014-McKenzie-JACS-136-7833, 2017-McKenzie-JCP-146-244903, 2017-McFadden-CM-29-10187, 2019-McFadden-PRB-99-125201, 2020-McFadden-PRB-102-235206}, and the study in soft condensed matter applications (see e.g.,~\cite{2014-McGee-JPCS-551-012039, 2015-McKenzie-SM-11-1755, 2018-McKenzie-SM-14-7324}) where the previously measured signal was completely wiped out below \SI{\sim 24}{\milli\tesla}, but access to fields an order of magnitude higher may ameliorate this difficulty.

Recent measurements on \ch{Nb} samples indicate vortex penetration at applied fields of \SI{98}{\milli\tesla} when \ch{Nb} samples are warmed close to \gls{tc}.
The mixed superconducting state is therefore also accessible and could be of interest for fundamental studies related to the superconducting vortex motion and dissipation~\cite{Vortex_Eley2021}.
\gls{bnmr} studies of the vortex lattice have been carried out in other superconductor materials such as \ch{YBa_2Cu_3O_{7-$\delta$}}\cite{vortex_YCBO_Saadaoui2009,vortex_YCBO_Saadaoui2009a} and \ch{NbSe2}\cite{vortex_NbSe2_Salman2007} using \gls{bnmr} spectrometer (with out-of-plane applied fields).
Further studies for \gls{srf} applications will benefit from lower sample temperatures in order to maintain the Meissner state at higher applied magnetic fields.
Installation of a procured cryocooler is ongoing and measurements of superconducting samples in this regime are expected in the near future.

		\section{
			Summary
			\label{sec:summary}
		}
		The instrumentation at TRIUMF's \gls{bnmr} facility allows for the depth-dependent characterization of the local magnetic field near the surface of a sample. Both the infrastructure and instrumentation have been upgraded with a new beamline extension on the \gls{bnqr} leg, which has expanded previous capabilities from a maximum parallel magnetic field of \SI{24}{\milli\tesla} to \SI{200}{\milli\tesla}. This capability is targeted at testing \gls{srf} samples in a regime analogous to the magnetic field conditions in a \ch{Nb} cavity operating at the fundamental limit, but will be widely useful to other condensed matter research. This capability is unique in the world and we anticipate additional use for other material science investigations.
	
		\begin{acknowledgments}
		We thank:
		L.~Merminga for involvement in the early stages of the project;
		M.~H.~Dehn for assistance during the beamline assembly;
		M.~Cervantes for sample preparation;
		D.~Lang, J.~Keir, R.~Abasalti, B.~Hitti, B.~Smith, D.~Vyas, T.~Au, J.~Chow, T.~Hruskovec, N.~Muller, and M.~Marchetto for providing excellent technical support.
		The hardware was funded through a Research Tools and Infrastructure grant from NSERC. 
	\end{acknowledgments}

\section*{Author Declarations}
\subsection*{Conflict of Interest}
The authors have no conflicts to disclose.

\subsection*{Author Contributions}
\textbf{Edward~Thoeng:} Data Curation (equal); Formal Analysis (equal); Investigation (equal); Validation (equal); Visualization (equal); Writing --- Original Draft Preparation (lead); Writing --- Review \& Editing (equal).
\textbf{Ryan~M.~L.~McFadden:} Data Curation (equal); Formal Analysis (equal); Investigation (supporting); Visualization (equal); Writing --- Review \& Editing (equal).
\textbf{Suresh~Saminathan:} Investigation (supporting); Methodology (equal); Validation (equal); Visualization (supporting).
\textbf{Gerald~D.~Morris:} Conceptualization (equal); Investigation (equal); Project Administration (supporting); Resources (equal); Supervision (supporting); Validation (equal).
\textbf{Philipp~Kolb:} Investigation (equal); Project Administration (supporting); Supervision (supporting) Validation (equal).
\textbf{Ben~Matheson:} Methodology (equal); Visualization (supporting).
\textbf{Md~Asaduzzaman:} Investigation (supporting).
\textbf{Richard~Baartman:} Conceptualization (equal); Supervision (supporting).
\textbf{Sarah~R.~Dunsiger:} Investigation (supporting).
\textbf{Derek~Fujimoto:} Formal Analysis (supporting); Investigation (supporting).
\textbf{Tobias~Junginger:} Formal Analysis (supporting); Investigation (supporting); Project Administration (supporting); Supervision (supporting).
\textbf{Victoria~L.~Karner:} Investigation (supporting).
\textbf{Spencer~Kiy:} Data Curation (equal); Investigation (supporting).
\textbf{Ruohong~Li:} Resources (equal).
\textbf{Monika~Stachura:} Investigation (supporting).
\textbf{John~O.~Ticknor:} Investigation (supporting).
\textbf{Robert~F.~Kiefl:} Conceptualization (equal); Funding Acquisition (supporting).
\textbf{W.~Andrew~MacFarlane:} Investigation (supporting); Supervision (supporting).
\textbf{Robert~E.~Laxdal:} Conceptualization (equal); Formal Analysis (supporting); Funding Acquisition (lead); Project Administration (lead); Supervision (lead); Writing --- Review \& Editing (equal).

\section*{Data Availability Statement}

The data that support the findings of this study are available from the corresponding author upon reasonable request. Raw data of the \gls{bnmr} experiments are available for download from: \url{https://cmms.triumf.ca}/


\begin{thebibliography}{58}%
		\makeatletter
		\providecommand \@ifxundefined [1]{%
			\@ifx{#1\undefined}
		}%
		\providecommand \@ifnum [1]{%
			\ifnum #1\expandafter \@firstoftwo
			\else \expandafter \@secondoftwo
			\fi
		}%
		\providecommand \@ifx [1]{%
			\ifx #1\expandafter \@firstoftwo
			\else \expandafter \@secondoftwo
			\fi
		}%
		\providecommand \natexlab [1]{#1}%
		\providecommand \enquote  [1]{``#1''}%
		\providecommand \bibnamefont  [1]{#1}%
		\providecommand \bibfnamefont [1]{#1}%
		\providecommand \citenamefont [1]{#1}%
		\providecommand \href@noop [0]{\@secondoftwo}%
		\providecommand \href [0]{\begingroup \@sanitize@url \@href}%
		\providecommand \@href[1]{\@@startlink{#1}\@@href}%
		\providecommand \@@href[1]{\endgroup#1\@@endlink}%
		\providecommand \@sanitize@url [0]{\catcode `\\12\catcode `\$12\catcode
			`\&12\catcode `\#12\catcode `\^12\catcode `\_12\catcode `\%12\relax}%
		\providecommand \@@startlink[1]{}%
		\providecommand \@@endlink[0]{}%
		\providecommand \url  [0]{\begingroup\@sanitize@url \@url }%
		\providecommand \@url [1]{\endgroup\@href {#1}{\urlprefix }}%
		\providecommand \urlprefix  [0]{URL }%
		\providecommand \Eprint [0]{\href }%
		\providecommand \doibase [0]{https://doi.org/}%
		\providecommand \selectlanguage [0]{\@gobble}%
		\providecommand \bibinfo  [0]{\@secondoftwo}%
		\providecommand \bibfield  [0]{\@secondoftwo}%
		\providecommand \translation [1]{[#1]}%
		\providecommand \BibitemOpen [0]{}%
		\providecommand \bibitemStop [0]{}%
		\providecommand \bibitemNoStop [0]{.\EOS\space}%
		\providecommand \EOS [0]{\spacefactor3000\relax}%
		\providecommand \BibitemShut  [1]{\csname bibitem#1\endcsname}%
		\let\auto@bib@innerbib\@empty
		\bibitem [{\citenamefont {{Padamsee}}(2019)}]{Padamsee2019}%
		\BibitemOpen
		\bibfield  {author} {\bibinfo {author} {\bibfnamefont {H.}~\bibnamefont
				{{Padamsee}}},\ }\bibfield  {title} {\enquote {\bibinfo {title} {{Future
						Prospects of Superconducting RF for Accelerator Applications}},}\ }\href
		{https://doi.org/10.1142/S1793626819300081} {\bibfield  {journal} {\bibinfo
				{journal} {Rev. Accel Sci. Technol.}\ }\textbf {\bibinfo {volume} {10}},\
			\bibinfo {pages} {125--156} (\bibinfo {year} {2019})}\BibitemShut {NoStop}%
		\bibitem [{\citenamefont {Padamsee}(2017)}]{Padamsee2017}%
		\BibitemOpen
		\bibfield  {author} {\bibinfo {author} {\bibfnamefont {H.}~\bibnamefont
				{Padamsee}},\ }\bibfield  {title} {\enquote {\bibinfo {title} {50 years of
					success for {SRF} accelerators{\textemdash}a review},}\ }\href
		{https://doi.org/10.1088/1361-6668/aa6376} {\bibfield  {journal} {\bibinfo
				{journal} {Supercond. Sci. Technol.}\ }\textbf {\bibinfo {volume} {30}},\
			\bibinfo {pages} {053003} (\bibinfo {year} {2017})}\BibitemShut {NoStop}%
		\bibitem [{\citenamefont {Gurevich}(2017)}]{Gurevich2017}%
		\BibitemOpen
		\bibfield  {author} {\bibinfo {author} {\bibfnamefont {A.}~\bibnamefont
				{Gurevich}},\ }\bibfield  {title} {\enquote {\bibinfo {title} {Theory of {RF}
					superconductivity for resonant cavities},}\ }\href
		{https://doi.org/10.1088/1361-6668/30/3/034004} {\bibfield  {journal}
			{\bibinfo  {journal} {Supercond. Sci. Technol.}\ }\textbf {\bibinfo {volume}
				{30}},\ \bibinfo {pages} {034004} (\bibinfo {year} {2017})}\BibitemShut
		{NoStop}%
		\bibitem [{\citenamefont {Ciovati}(2004)}]{Ciovati2004}%
		\BibitemOpen
		\bibfield  {author} {\bibinfo {author} {\bibfnamefont {G.}~\bibnamefont
				{Ciovati}},\ }\bibfield  {title} {\enquote {\bibinfo {title} {Effect of
					low-temperature baking on the radio-frequency properties of niobium
					superconducting cavities for particle accelerators},}\ }\href
		{https://doi.org/10.1063/1.1767295} {\bibfield  {journal} {\bibinfo
				{journal} {J. Appl. Phys.}\ }\textbf {\bibinfo {volume} {96}},\ \bibinfo
			{pages} {1591--1600} (\bibinfo {year} {2004})}\BibitemShut {NoStop}%
		\bibitem [{\citenamefont {He}\ \emph {et~al.}(2021)\citenamefont {He},
			\citenamefont {Pan}, \citenamefont {Sha}, \citenamefont {Zhai}, \citenamefont
			{Mi}, \citenamefont {Dai}, \citenamefont {Jin}, \citenamefont {Zhang},
			\citenamefont {Dong}, \citenamefont {Liu}, \citenamefont {Zhao},
			\citenamefont {Ge}, \citenamefont {Zhao}, \citenamefont {Mu}, \citenamefont
			{Du}, \citenamefont {Sun}, \citenamefont {Zhang}, \citenamefont {Yang},\ and\
			\citenamefont {Zheng}}]{He2021}%
		\BibitemOpen
		\bibfield  {author} {\bibinfo {author} {\bibfnamefont {F.}~\bibnamefont
				{He}}, \bibinfo {author} {\bibfnamefont {W.}~\bibnamefont {Pan}}, \bibinfo
			{author} {\bibfnamefont {P.}~\bibnamefont {Sha}}, \bibinfo {author}
			{\bibfnamefont {J.}~\bibnamefont {Zhai}}, \bibinfo {author} {\bibfnamefont
				{Z.}~\bibnamefont {Mi}}, \bibinfo {author} {\bibfnamefont {X.}~\bibnamefont
				{Dai}}, \bibinfo {author} {\bibfnamefont {S.}~\bibnamefont {Jin}}, \bibinfo
			{author} {\bibfnamefont {Z.}~\bibnamefont {Zhang}}, \bibinfo {author}
			{\bibfnamefont {C.}~\bibnamefont {Dong}}, \bibinfo {author} {\bibfnamefont
				{B.}~\bibnamefont {Liu}}, \bibinfo {author} {\bibfnamefont {H.}~\bibnamefont
				{Zhao}}, \bibinfo {author} {\bibfnamefont {R.}~\bibnamefont {Ge}}, \bibinfo
			{author} {\bibfnamefont {J.}~\bibnamefont {Zhao}}, \bibinfo {author}
			{\bibfnamefont {Z.}~\bibnamefont {Mu}}, \bibinfo {author} {\bibfnamefont
				{L.}~\bibnamefont {Du}}, \bibinfo {author} {\bibfnamefont {L.}~\bibnamefont
				{Sun}}, \bibinfo {author} {\bibfnamefont {L.}~\bibnamefont {Zhang}}, \bibinfo
			{author} {\bibfnamefont {C.}~\bibnamefont {Yang}},\ and\ \bibinfo {author}
			{\bibfnamefont {X.}~\bibnamefont {Zheng}},\ }\bibfield  {title} {\enquote
			{\bibinfo {title} {Medium-temperature furnace baking of 1.3 {GHz} 9-cell
					superconducting cavities at {IHEP}},}\ }\href
		{https://doi.org/10.1088/1361-6668/ac1657} {\bibfield  {journal} {\bibinfo
				{journal} {Supercond. Sci. Technol.}\ }\textbf {\bibinfo {volume} {34}},\
			\bibinfo {pages} {095005} (\bibinfo {year} {2021})}\BibitemShut {NoStop}%
		\bibitem [{\citenamefont {Ito}\ \emph {et~al.}(2021)\citenamefont {Ito},
			\citenamefont {Araki}, \citenamefont {Takahashi},\ and\ \citenamefont
			{Umemori}}]{Ito2021}%
		\BibitemOpen
		\bibfield  {author} {\bibinfo {author} {\bibfnamefont {H.}~\bibnamefont
				{Ito}}, \bibinfo {author} {\bibfnamefont {H.}~\bibnamefont {Araki}}, \bibinfo
			{author} {\bibfnamefont {K.}~\bibnamefont {Takahashi}},\ and\ \bibinfo
			{author} {\bibfnamefont {K.}~\bibnamefont {Umemori}},\ }\bibfield  {title}
		{\enquote {\bibinfo {title} {{Influence of furnace baking on Q–E behavior
						of superconducting accelerating cavities}},}\ }\href
		{https://doi.org/10.1093/ptep/ptab056} {\bibfield  {journal} {\bibinfo
				{journal} {Prog. Theor. Exp. Phys.}\ }\textbf {\bibinfo {volume} {2021}}
			(\bibinfo {year} {2021}),\ 10.1093/ptep/ptab056},\ \bibinfo {note}
		{071G01}\BibitemShut {NoStop}%
		\bibitem [{\citenamefont {Lechner}\ \emph {et~al.}(2021)\citenamefont
			{Lechner}, \citenamefont {Angle}, \citenamefont {Stevie}, \citenamefont
			{Kelley}, \citenamefont {Reece},\ and\ \citenamefont
			{Palczewski}}]{Lechner2021}%
		\BibitemOpen
		\bibfield  {author} {\bibinfo {author} {\bibfnamefont {E.~M.}\ \bibnamefont
				{Lechner}}, \bibinfo {author} {\bibfnamefont {J.~W.}\ \bibnamefont {Angle}},
			\bibinfo {author} {\bibfnamefont {F.~A.}\ \bibnamefont {Stevie}}, \bibinfo
			{author} {\bibfnamefont {M.~J.}\ \bibnamefont {Kelley}}, \bibinfo {author}
			{\bibfnamefont {C.~E.}\ \bibnamefont {Reece}},\ and\ \bibinfo {author}
			{\bibfnamefont {A.~D.}\ \bibnamefont {Palczewski}},\ }\bibfield  {title}
		{\enquote {\bibinfo {title} {Rf surface resistance tuning of superconducting
					niobium via thermal diffusion of native oxide},}\ }\href
		{https://doi.org/10.1063/5.0059464} {\bibfield  {journal} {\bibinfo
				{journal} {Appl. Phys. Lett.}\ }\textbf {\bibinfo {volume} {119}},\ \bibinfo
			{pages} {082601} (\bibinfo {year} {2021})}\BibitemShut {NoStop}%
		\bibitem [{\citenamefont {Grassellino}\ \emph
			{et~al.}(2013{\natexlab{a}})\citenamefont {Grassellino}, \citenamefont
			{Romanenko}, \citenamefont {Sergatskov}, \citenamefont {Melnychuk},
			\citenamefont {Trenikhina}, \citenamefont {Crawford}, \citenamefont {Rowe},
			\citenamefont {Wong}, \citenamefont {Khabiboulline},\ and\ \citenamefont
			{Barkov}}]{Grassellino2013}%
		\BibitemOpen
		\bibfield  {author} {\bibinfo {author} {\bibfnamefont {A.}~\bibnamefont
				{Grassellino}}, \bibinfo {author} {\bibfnamefont {A.}~\bibnamefont
				{Romanenko}}, \bibinfo {author} {\bibfnamefont {D.}~\bibnamefont
				{Sergatskov}}, \bibinfo {author} {\bibfnamefont {O.}~\bibnamefont
				{Melnychuk}}, \bibinfo {author} {\bibfnamefont {Y.}~\bibnamefont
				{Trenikhina}}, \bibinfo {author} {\bibfnamefont {A.}~\bibnamefont
				{Crawford}}, \bibinfo {author} {\bibfnamefont {A.}~\bibnamefont {Rowe}},
			\bibinfo {author} {\bibfnamefont {M.}~\bibnamefont {Wong}}, \bibinfo {author}
			{\bibfnamefont {T.}~\bibnamefont {Khabiboulline}},\ and\ \bibinfo {author}
			{\bibfnamefont {F.}~\bibnamefont {Barkov}},\ }\bibfield  {title} {\enquote
			{\bibinfo {title} {Nitrogen and argon doping of niobium for superconducting
					radio frequency cavities: a pathway to highly efficient accelerating
					structures},}\ }\href {https://doi.org/10.1088/0953-2048/26/10/102001}
		{\bibfield  {journal} {\bibinfo  {journal} {Supercond. Sci. Technol.}\
			}\textbf {\bibinfo {volume} {26}},\ \bibinfo {pages} {102001} (\bibinfo
			{year} {2013}{\natexlab{a}})}\BibitemShut {NoStop}%
		\bibitem [{\citenamefont {Grassellino}\ \emph {et~al.}(2017)\citenamefont
			{Grassellino}, \citenamefont {Romanenko}, \citenamefont {Trenikhina},
			\citenamefont {Checchin}, \citenamefont {Martinello}, \citenamefont
			{Melnychuk}, \citenamefont {Chandrasekaran}, \citenamefont {Sergatskov},
			\citenamefont {Posen}, \citenamefont {Crawford}, \citenamefont {Aderhold},\
			and\ \citenamefont {Bice}}]{Grassellino2017}%
		\BibitemOpen
		\bibfield  {author} {\bibinfo {author} {\bibfnamefont {A.}~\bibnamefont
				{Grassellino}}, \bibinfo {author} {\bibfnamefont {A.}~\bibnamefont
				{Romanenko}}, \bibinfo {author} {\bibfnamefont {Y.}~\bibnamefont
				{Trenikhina}}, \bibinfo {author} {\bibfnamefont {M.}~\bibnamefont
				{Checchin}}, \bibinfo {author} {\bibfnamefont {M.}~\bibnamefont
				{Martinello}}, \bibinfo {author} {\bibfnamefont {O.~S.}\ \bibnamefont
				{Melnychuk}}, \bibinfo {author} {\bibfnamefont {S.}~\bibnamefont
				{Chandrasekaran}}, \bibinfo {author} {\bibfnamefont {D.~A.}\ \bibnamefont
				{Sergatskov}}, \bibinfo {author} {\bibfnamefont {S.}~\bibnamefont {Posen}},
			\bibinfo {author} {\bibfnamefont {A.~C.}\ \bibnamefont {Crawford}}, \bibinfo
			{author} {\bibfnamefont {S.}~\bibnamefont {Aderhold}},\ and\ \bibinfo
			{author} {\bibfnamefont {D.}~\bibnamefont {Bice}},\ }\bibfield  {title}
		{\enquote {\bibinfo {title} {Unprecedented quality factors at accelerating
					gradients up to 45 {MVm}$^{-1}$ in niobium superconducting resonators via low
					temperature nitrogen infusion},}\ }\href
		{https://doi.org/10.1088/1361-6668/aa7afe} {\bibfield  {journal} {\bibinfo
				{journal} {Supercond. Sci. Technol.}\ }\textbf {\bibinfo {volume} {30}},\
			\bibinfo {pages} {094004} (\bibinfo {year} {2017})}\BibitemShut {NoStop}%
		\bibitem [{\citenamefont {Kubo}(2016)}]{Kubo2016}%
		\BibitemOpen
		\bibfield  {author} {\bibinfo {author} {\bibfnamefont {T.}~\bibnamefont
				{Kubo}},\ }\bibfield  {title} {\enquote {\bibinfo {title} {Multilayer coating
					for higher accelerating fields in superconducting radio-frequency cavities: a
					review of theoretical aspects},}\ }\href
		{https://doi.org/10.1088/1361-6668/30/2/023001} {\bibfield  {journal}
			{\bibinfo  {journal} {Supercond. Sci. Technol.}\ }\textbf {\bibinfo {volume}
				{30}},\ \bibinfo {pages} {023001} (\bibinfo {year} {2016})}\BibitemShut
		{NoStop}%
		\bibitem [{\citenamefont {Gurevich}(2006)}]{Gurevich2006}%
		\BibitemOpen
		\bibfield  {author} {\bibinfo {author} {\bibfnamefont {A.}~\bibnamefont
				{Gurevich}},\ }\bibfield  {title} {\enquote {\bibinfo {title} {Enhancement of
					rf breakdown field of superconductors by multilayer coating},}\ }\href
		{https://doi.org/10.1063/1.2162264} {\bibfield  {journal} {\bibinfo
				{journal} {Appl. Phys. Lett.}\ }\textbf {\bibinfo {volume} {88}},\ \bibinfo
			{pages} {012511} (\bibinfo {year} {2006})}\BibitemShut {NoStop}%
		\bibitem [{\citenamefont {Gurevich}(2015)}]{Gurevich2015}%
		\BibitemOpen
		\bibfield  {author} {\bibinfo {author} {\bibfnamefont {A.}~\bibnamefont
				{Gurevich}},\ }\bibfield  {title} {\enquote {\bibinfo {title} {Maximum
					screening fields of superconducting multilayer structures},}\ }\href
		{https://doi.org/10.1063/1.4905711} {\bibfield  {journal} {\bibinfo
				{journal} {AIP Advances}\ }\textbf {\bibinfo {volume} {5}},\ \bibinfo {pages}
			{017112} (\bibinfo {year} {2015})}\BibitemShut {NoStop}%
		\bibitem [{\citenamefont {Posen}\ \emph {et~al.}(2015)\citenamefont {Posen},
			\citenamefont {Transtrum}, \citenamefont {Catelani}, \citenamefont {Liepe},\
			and\ \citenamefont {Sethna}}]{Posen2015}%
		\BibitemOpen
		\bibfield  {author} {\bibinfo {author} {\bibfnamefont {S.}~\bibnamefont
				{Posen}}, \bibinfo {author} {\bibfnamefont {M.~K.}\ \bibnamefont
				{Transtrum}}, \bibinfo {author} {\bibfnamefont {G.}~\bibnamefont {Catelani}},
			\bibinfo {author} {\bibfnamefont {M.~U.}\ \bibnamefont {Liepe}},\ and\
			\bibinfo {author} {\bibfnamefont {J.~P.}\ \bibnamefont {Sethna}},\ }\bibfield
		{title} {\enquote {\bibinfo {title} {Shielding {S}uperconductors with {T}hin
					{F}ilms as {A}pplied to rf {C}avities for {P}article {A}ccelerators},}\
		}\href {https://doi.org/10.1103/PhysRevApplied.4.044019} {\bibfield
			{journal} {\bibinfo  {journal} {Phys. Rev. Applied}\ }\textbf {\bibinfo
				{volume} {4}},\ \bibinfo {pages} {044019} (\bibinfo {year}
			{2015})}\BibitemShut {NoStop}%
		\bibitem [{\citenamefont {Checchin}\ \emph {et~al.}(2016)\citenamefont
			{Checchin}, \citenamefont {Grassellino}, \citenamefont {Martinello},
			\citenamefont {Posen}, \citenamefont {Romanenko},\ and\ \citenamefont
			{Zasadzinski}}]{Checchin_GL_theory}%
		\BibitemOpen
		\bibfield  {author} {\bibinfo {author} {\bibfnamefont {M.}~\bibnamefont
				{Checchin}}, \bibinfo {author} {\bibfnamefont {A.}~\bibnamefont
				{Grassellino}}, \bibinfo {author} {\bibfnamefont {M.}~\bibnamefont
				{Martinello}}, \bibinfo {author} {\bibfnamefont {S.}~\bibnamefont {Posen}},
			\bibinfo {author} {\bibfnamefont {A.}~\bibnamefont {Romanenko}},\ and\
			\bibinfo {author} {\bibfnamefont {J.}~\bibnamefont {Zasadzinski}},\
		}\bibfield  {title} {\enquote {\bibinfo {title} {{U}ltimate {G}radient
					{L}imitation in {N}iobium {S}uperconducting {A}ccelerating {C}avities},}\
		}in\ \href {https://doi.org/doi:10.18429/JACoW-IPAC2016-WEPMR002} {\emph
			{\bibinfo {booktitle} {Proc. of International Particle Accelerator Conference
					(IPAC'16), Busan, Korea, May 8-13, 2016}}},\ \bibinfo {series and number}
		{\bibinfo {series} {International Particle Accelerator Conference}\
			No.~\bibinfo {number} {7}}\ (\bibinfo  {publisher} {JACoW},\ \bibinfo
		{address} {Geneva, Switzerland},\ \bibinfo {year} {2016})\ pp.\ \bibinfo
		{pages} {2254--2257},\ \bibinfo {note}
		{doi:10.18429/JACoW-IPAC2016-WEPMR002}\BibitemShut {NoStop}%
		\bibitem [{\citenamefont {Junginger}\ \emph {et~al.}(2018)\citenamefont
			{Junginger}, \citenamefont {Abidi}, \citenamefont {Maffett}, \citenamefont
			{Buck}, \citenamefont {Dehn}, \citenamefont {Gheidi}, \citenamefont {Kiefl},
			\citenamefont {Kolb}, \citenamefont {Storey}, \citenamefont {Thoeng},
			\citenamefont {Wasserman},\ and\ \citenamefont {Laxdal}}]{Junginger2018}%
		\BibitemOpen
		\bibfield  {author} {\bibinfo {author} {\bibfnamefont {T.}~\bibnamefont
				{Junginger}}, \bibinfo {author} {\bibfnamefont {S.~H.}\ \bibnamefont
				{Abidi}}, \bibinfo {author} {\bibfnamefont {R.~D.}\ \bibnamefont {Maffett}},
			\bibinfo {author} {\bibfnamefont {T.}~\bibnamefont {Buck}}, \bibinfo {author}
			{\bibfnamefont {M.~H.}\ \bibnamefont {Dehn}}, \bibinfo {author}
			{\bibfnamefont {S.}~\bibnamefont {Gheidi}}, \bibinfo {author} {\bibfnamefont
				{R.}~\bibnamefont {Kiefl}}, \bibinfo {author} {\bibfnamefont
				{P.}~\bibnamefont {Kolb}}, \bibinfo {author} {\bibfnamefont {D.}~\bibnamefont
				{Storey}}, \bibinfo {author} {\bibfnamefont {E.}~\bibnamefont {Thoeng}},
			\bibinfo {author} {\bibfnamefont {W.}~\bibnamefont {Wasserman}},\ and\
			\bibinfo {author} {\bibfnamefont {R.~E.}\ \bibnamefont {Laxdal}},\ }\bibfield
		{title} {\enquote {\bibinfo {title} {Field of first magnetic flux entry and
					pinning strength of superconductors for rf application measured with muon
					spin rotation},}\ }\href
		{https://doi.org/10.1103/PhysRevAccelBeams.21.032002} {\bibfield  {journal}
			{\bibinfo  {journal} {Phys. Rev. Accel. Beams}\ }\textbf {\bibinfo {volume}
				{21}},\ \bibinfo {pages} {032002} (\bibinfo {year} {2018})}\BibitemShut
		{NoStop}%
		\bibitem [{\citenamefont {Grassellino}\ \emph
			{et~al.}(2013{\natexlab{b}})\citenamefont {Grassellino}, \citenamefont
			{Beard}, \citenamefont {Kolb}, \citenamefont {Laxdal}, \citenamefont
			{Lockyer}, \citenamefont {Longuevergne},\ and\ \citenamefont
			{Sonier}}]{Grassellino_muSR_TRIUMForig}%
		\BibitemOpen
		\bibfield  {author} {\bibinfo {author} {\bibfnamefont {A.}~\bibnamefont
				{Grassellino}}, \bibinfo {author} {\bibfnamefont {C.}~\bibnamefont {Beard}},
			\bibinfo {author} {\bibfnamefont {P.}~\bibnamefont {Kolb}}, \bibinfo {author}
			{\bibfnamefont {R.}~\bibnamefont {Laxdal}}, \bibinfo {author} {\bibfnamefont
				{N.~S.}\ \bibnamefont {Lockyer}}, \bibinfo {author} {\bibfnamefont
				{D.}~\bibnamefont {Longuevergne}},\ and\ \bibinfo {author} {\bibfnamefont
				{J.~E.}\ \bibnamefont {Sonier}},\ }\bibfield  {title} {\enquote {\bibinfo
				{title} {Muon spin rotation studies of niobium for superconducting rf
					applications},}\ }\href {https://doi.org/10.1103/PhysRevSTAB.16.062002}
		{\bibfield  {journal} {\bibinfo  {journal} {Phys. Rev. ST Accel. Beams}\
			}\textbf {\bibinfo {volume} {16}},\ \bibinfo {pages} {062002} (\bibinfo
			{year} {2013}{\natexlab{b}})}\BibitemShut {NoStop}%
		\bibitem [{\citenamefont {Laxdal}\ \emph {et~al.}(2015)\citenamefont {Laxdal},
			\citenamefont {Abidi}, \citenamefont {Buck}, \citenamefont {Junginger},
			\citenamefont {Kiefl}, \citenamefont {Kolb}, \citenamefont {Ma},
			\citenamefont {Yang},\ and\ \citenamefont {Yao}}]{Gheidi_SRF2015}%
		\BibitemOpen
		\bibfield  {author} {\bibinfo {author} {\bibfnamefont {R.}~\bibnamefont
				{Laxdal}}, \bibinfo {author} {\bibfnamefont {S.}~\bibnamefont {Abidi}},
			\bibinfo {author} {\bibfnamefont {T.}~\bibnamefont {Buck}}, \bibinfo {author}
			{\bibfnamefont {T.}~\bibnamefont {Junginger}}, \bibinfo {author}
			{\bibfnamefont {R.}~\bibnamefont {Kiefl}}, \bibinfo {author} {\bibfnamefont
				{P.}~\bibnamefont {Kolb}}, \bibinfo {author} {\bibfnamefont {Y.}~\bibnamefont
				{Ma}}, \bibinfo {author} {\bibfnamefont {L.}~\bibnamefont {Yang}},\ and\
			\bibinfo {author} {\bibfnamefont {Z.}~\bibnamefont {Yao}},\ }\bibfield
		{title} {\enquote {\bibinfo {title} {{Characterization of SRF Materials at
						the TRIUMF muSR Facility}},}\ }in\ \href
		{https://doi.org/10.18429/JACoW-SRF2015-MOPB050} {\emph {\bibinfo {booktitle}
				{{17th International Conference on RF Superconductivity}}}}\ (\bibinfo {year}
		{2015})\ p.\ \bibinfo {pages} {MOPB050}\BibitemShut {NoStop}%
		\bibitem [{\citenamefont {Salman}\ \emph {et~al.}(2012)\citenamefont {Salman},
			\citenamefont {Prokscha}, \citenamefont {Keller}, \citenamefont {Morenzoni},
			\citenamefont {Saadaoui}, \citenamefont {Sedlak}, \citenamefont {Shiroka},
			\citenamefont {Sidorov}, \citenamefont {Suter}, \citenamefont {Vrankovic},\
			and\ \citenamefont {Weber}}]{Salman2012}%
		\BibitemOpen
		\bibfield  {author} {\bibinfo {author} {\bibfnamefont {Z.}~\bibnamefont
				{Salman}}, \bibinfo {author} {\bibfnamefont {T.}~\bibnamefont {Prokscha}},
			\bibinfo {author} {\bibfnamefont {P.}~\bibnamefont {Keller}}, \bibinfo
			{author} {\bibfnamefont {E.}~\bibnamefont {Morenzoni}}, \bibinfo {author}
			{\bibfnamefont {H.}~\bibnamefont {Saadaoui}}, \bibinfo {author}
			{\bibfnamefont {K.}~\bibnamefont {Sedlak}}, \bibinfo {author} {\bibfnamefont
				{T.}~\bibnamefont {Shiroka}}, \bibinfo {author} {\bibfnamefont
				{S.}~\bibnamefont {Sidorov}}, \bibinfo {author} {\bibfnamefont
				{A.}~\bibnamefont {Suter}}, \bibinfo {author} {\bibfnamefont
				{V.}~\bibnamefont {Vrankovic}},\ and\ \bibinfo {author} {\bibfnamefont
				{H.-P.}\ \bibnamefont {Weber}},\ }\bibfield  {title} {\enquote {\bibinfo
				{title} {Design and simulation of a spin rotator for longitudinal field
					measurements in the low energy muons spectrometer},}\ }\href
		{https://doi.org/https://doi.org/10.1016/j.phpro.2012.04.039} {\bibfield
			{journal} {\bibinfo  {journal} {Physics Procedia}\ }\textbf {\bibinfo
				{volume} {30}},\ \bibinfo {pages} {55--60} (\bibinfo {year} {2012})},\
		\bibinfo {note} {12th International Conference on Muon Spin Rotation,
			Relaxation and Resonance ({$\mu$}SR2011)}\BibitemShut {NoStop}%
		\bibitem [{\citenamefont {Prokscha}\ \emph {et~al.}(2008)\citenamefont
			{Prokscha}, \citenamefont {Morenzoni}, \citenamefont {Deiters}, \citenamefont
			{Foroughi}, \citenamefont {George}, \citenamefont {Kobler}, \citenamefont
			{Suter},\ and\ \citenamefont {Vrankovic}}]{2008-Prokscha-NIMA-595-317}%
		\BibitemOpen
		\bibfield  {author} {\bibinfo {author} {\bibfnamefont {T.}~\bibnamefont
				{Prokscha}}, \bibinfo {author} {\bibfnamefont {E.}~\bibnamefont {Morenzoni}},
			\bibinfo {author} {\bibfnamefont {K.}~\bibnamefont {Deiters}}, \bibinfo
			{author} {\bibfnamefont {F.}~\bibnamefont {Foroughi}}, \bibinfo {author}
			{\bibfnamefont {D.}~\bibnamefont {George}}, \bibinfo {author} {\bibfnamefont
				{R.}~\bibnamefont {Kobler}}, \bibinfo {author} {\bibfnamefont
				{A.}~\bibnamefont {Suter}},\ and\ \bibinfo {author} {\bibfnamefont
				{V.}~\bibnamefont {Vrankovic}},\ }\bibfield  {title} {\enquote {\bibinfo
				{title} {The new {{$\mu$}E4} beam at {PSI}: A hybrid-type large acceptance
					channel for the generation of a high intensity surface-muon beam},}\ }\href
		{https://doi.org/10.1016/j.nima.2008.07.081} {\bibfield  {journal} {\bibinfo
				{journal} {Nucl. Instrum. Methods Phys. Res., Sect. A}\ }\textbf {\bibinfo
				{volume} {595}},\ \bibinfo {pages} {317--331} (\bibinfo {year}
			{2008})}\BibitemShut {NoStop}%
		\bibitem [{\citenamefont {Morenzoni}\ \emph {et~al.}(2000)\citenamefont
			{Morenzoni}, \citenamefont {Glückler}, \citenamefont {Prokscha},
			\citenamefont {Weber}, \citenamefont {Forgan}, \citenamefont {Jackson},
			\citenamefont {Luetkens}, \citenamefont {Niedermayer}, \citenamefont
			{Pleines}, \citenamefont {Birke}, \citenamefont {Hofer}, \citenamefont
			{Litterst}, \citenamefont {Riseman},\ and\ \citenamefont
			{Schatz}}]{Morenzoni2000}%
		\BibitemOpen
		\bibfield  {author} {\bibinfo {author} {\bibfnamefont {E.}~\bibnamefont
				{Morenzoni}}, \bibinfo {author} {\bibfnamefont {H.}~\bibnamefont
				{Glückler}}, \bibinfo {author} {\bibfnamefont {T.}~\bibnamefont {Prokscha}},
			\bibinfo {author} {\bibfnamefont {H.}~\bibnamefont {Weber}}, \bibinfo
			{author} {\bibfnamefont {E.}~\bibnamefont {Forgan}}, \bibinfo {author}
			{\bibfnamefont {T.}~\bibnamefont {Jackson}}, \bibinfo {author} {\bibfnamefont
				{H.}~\bibnamefont {Luetkens}}, \bibinfo {author} {\bibfnamefont
				{C.}~\bibnamefont {Niedermayer}}, \bibinfo {author} {\bibfnamefont
				{M.}~\bibnamefont {Pleines}}, \bibinfo {author} {\bibfnamefont
				{M.}~\bibnamefont {Birke}}, \bibinfo {author} {\bibfnamefont
				{A.}~\bibnamefont {Hofer}}, \bibinfo {author} {\bibfnamefont
				{J.}~\bibnamefont {Litterst}}, \bibinfo {author} {\bibfnamefont
				{T.}~\bibnamefont {Riseman}},\ and\ \bibinfo {author} {\bibfnamefont
				{G.}~\bibnamefont {Schatz}},\ }\bibfield  {title} {\enquote {\bibinfo {title}
				{Low-energy {$\mu$}{SR} at {PSI}: present and future},}\ }\href
		{https://doi.org/https://doi.org/10.1016/S0921-4526(00)00303-3} {\bibfield
			{journal} {\bibinfo  {journal} {Physica B}\ }\textbf {\bibinfo {volume}
				{289-290}},\ \bibinfo {pages} {653--657} (\bibinfo {year}
			{2000})}\BibitemShut {NoStop}%
		\bibitem [{\citenamefont {MacFarlane}(2022)}]{MacFarlane_ZPC2021}%
		\BibitemOpen
		\bibfield  {author} {\bibinfo {author} {\bibfnamefont {W.~A.}\ \bibnamefont
				{MacFarlane}},\ }\bibfield  {title} {\enquote {\bibinfo {title} {Status and
					progress of ion-implanted {$\beta$}{NMR} at triumf},}\ }\href
		{https://doi.org/doi:10.1515/zpch-2021-3154} {\bibfield  {journal} {\bibinfo
				{journal} {Z. Phys. Chem.}\ }\textbf {\bibinfo {volume} {236}},\ \bibinfo
			{pages} {757--798} (\bibinfo {year} {2022})}\BibitemShut {NoStop}%
		\bibitem [{\citenamefont {MacFarlane}(2015)}]{2015-MacFarlane-SSNMR-68-1}%
		\BibitemOpen
		\bibfield  {author} {\bibinfo {author} {\bibfnamefont {W.~A.}\ \bibnamefont
				{MacFarlane}},\ }\bibfield  {title} {\enquote {\bibinfo {title}
				{Implanted-ion {{$\beta$}NMR}: A new probe for nanoscience},}\ }\href
		{https://doi.org/10.1016/j.ssnmr.2015.02.004} {\bibfield  {journal} {\bibinfo
				{journal} {Solid State Nucl. Magn. Reson.}\ }\textbf {\bibinfo {volume}
				{68--69}},\ \bibinfo {pages} {1--12} (\bibinfo {year} {2015})}\BibitemShut
		{NoStop}%
		\bibitem [{\citenamefont {Morris}(2014)}]{2014-Morris-HI-225-173}%
		\BibitemOpen
		\bibfield  {author} {\bibinfo {author} {\bibfnamefont {G.~D.}\ \bibnamefont
				{Morris}},\ }\bibfield  {title} {\enquote {\bibinfo {title}
				{{{$\beta$}-NMR}},}\ }\href {https://doi.org/10.1007/s10751-013-0894-6}
		{\bibfield  {journal} {\bibinfo  {journal} {Hyperfine Interact.}\ }\textbf
			{\bibinfo {volume} {225}},\ \bibinfo {pages} {173--182} (\bibinfo {year}
			{2014})}\BibitemShut {NoStop}%
		\bibitem [{\citenamefont {Kreitzman}\ and\ \citenamefont
			{Morris}(2018)}]{2018-Kreitzman-JPSCP-21-011056}%
		\BibitemOpen
		\bibfield  {author} {\bibinfo {author} {\bibfnamefont {S.~R.}\ \bibnamefont
				{Kreitzman}}\ and\ \bibinfo {author} {\bibfnamefont {G.~D.}\ \bibnamefont
				{Morris}},\ }\bibfield  {title} {\enquote {\bibinfo {title} {{TRIUMF} {MuSR}
					and {{$\beta$}NMR} research facilities},}\ }\href
		{https://doi.org/10.7566/JPSCP.21.011056} {\bibfield  {journal} {\bibinfo
				{journal} {JPS Conf. Proc.}\ }\textbf {\bibinfo {volume} {21}},\ \bibinfo
			{pages} {011056} (\bibinfo {year} {2018})}\BibitemShut {NoStop}%
		\bibitem [{\citenamefont {Fl\'echard}\ \emph {et~al.}(2010)\citenamefont
			{Fl\'echard}, \citenamefont {Li\'enard}, \citenamefont {Naviliat-Cuncic},
			\citenamefont {Rodr\'{\i}guez}, \citenamefont {Alvarez}, \citenamefont {Ban},
			\citenamefont {Carniol}, \citenamefont {Etasse}, \citenamefont {Fontbonne},
			\citenamefont {Lallena},\ and\ \citenamefont {Praena}}]{Flechard2010}%
		\BibitemOpen
		\bibfield  {author} {\bibinfo {author} {\bibfnamefont {X.}~\bibnamefont
				{Fl\'echard}}, \bibinfo {author} {\bibfnamefont {E.}~\bibnamefont
				{Li\'enard}}, \bibinfo {author} {\bibfnamefont {O.}~\bibnamefont
				{Naviliat-Cuncic}}, \bibinfo {author} {\bibfnamefont {D.}~\bibnamefont
				{Rodr\'{\i}guez}}, \bibinfo {author} {\bibfnamefont {M.~A.~G.}\ \bibnamefont
				{Alvarez}}, \bibinfo {author} {\bibfnamefont {G.}~\bibnamefont {Ban}},
			\bibinfo {author} {\bibfnamefont {B.}~\bibnamefont {Carniol}}, \bibinfo
			{author} {\bibfnamefont {D.}~\bibnamefont {Etasse}}, \bibinfo {author}
			{\bibfnamefont {J.~M.}\ \bibnamefont {Fontbonne}}, \bibinfo {author}
			{\bibfnamefont {A.~M.}\ \bibnamefont {Lallena}},\ and\ \bibinfo {author}
			{\bibfnamefont {J.}~\bibnamefont {Praena}},\ }\bibfield  {title} {\enquote
			{\bibinfo {title} {Measurement of the $^{8}\mathrm{Li}$ half-life},}\ }\href
		{https://doi.org/10.1103/PhysRevC.82.027309} {\bibfield  {journal} {\bibinfo
				{journal} {Phys. Rev. C}\ }\textbf {\bibinfo {volume} {82}},\ \bibinfo
			{pages} {027309} (\bibinfo {year} {2010})}\BibitemShut {NoStop}%
		\bibitem [{\citenamefont {Dilling}\ and\ \citenamefont
			{Krücken}(2014)}]{Dilling_Hyperfine_ISAC}%
		\BibitemOpen
		\bibfield  {author} {\bibinfo {author} {\bibfnamefont {J.}~\bibnamefont
				{Dilling}}\ and\ \bibinfo {author} {\bibfnamefont {R.}~\bibnamefont
				{Krücken}},\ }\bibfield  {title} {\enquote {\bibinfo {title} {The
					experimental facilities at {ISAC}},}\ }\href
		{https://doi.org/10.1007/s10751-013-0886-6} {\bibfield  {journal} {\bibinfo
				{journal} {Hyperfine Interact.}\ }\textbf {\bibinfo {volume} {225}},\
			\bibinfo {pages} {111--114} (\bibinfo {year} {2014})}\BibitemShut {NoStop}%
		\bibitem [{\citenamefont {Levy}\ \emph {et~al.}(2014)\citenamefont {Levy},
			\citenamefont {Pearson}, \citenamefont {Kiefl}, \citenamefont {Man{\'e}},
			\citenamefont {Morris},\ and\ \citenamefont {Voss}}]{2014-Levy-HI-225-165}%
		\BibitemOpen
		\bibfield  {author} {\bibinfo {author} {\bibfnamefont {C.~D.~P.}\
				\bibnamefont {Levy}}, \bibinfo {author} {\bibfnamefont {M.~R.}\ \bibnamefont
				{Pearson}}, \bibinfo {author} {\bibfnamefont {R.~F.}\ \bibnamefont {Kiefl}},
			\bibinfo {author} {\bibfnamefont {E.}~\bibnamefont {Man{\'e}}}, \bibinfo
			{author} {\bibfnamefont {G.~D.}\ \bibnamefont {Morris}},\ and\ \bibinfo
			{author} {\bibfnamefont {A.}~\bibnamefont {Voss}},\ }\bibfield  {title}
		{\enquote {\bibinfo {title} {Laser polarization facility},}\ }\href
		{https://doi.org/10.1007/s10751-013-0896-4} {\bibfield  {journal} {\bibinfo
				{journal} {Hyperfine Interact.}\ }\textbf {\bibinfo {volume} {225}},\
			\bibinfo {pages} {165--172} (\bibinfo {year} {2014})}\BibitemShut {NoStop}%
		\bibitem [{\citenamefont {MacFarlane}\ \emph {et~al.}(2014)\citenamefont
			{MacFarlane}, \citenamefont {Levy}, \citenamefont {Pearson}, \citenamefont
			{Buck}, \citenamefont {Chow}, \citenamefont {Hariwal}, \citenamefont {Kiefl},
			\citenamefont {McGee}, \citenamefont {Morris},\ and\ \citenamefont
			{Wang}}]{MacFarlane2014_InitSpinPol}%
		\BibitemOpen
		\bibfield  {author} {\bibinfo {author} {\bibfnamefont {W.~A.}\ \bibnamefont
				{MacFarlane}}, \bibinfo {author} {\bibfnamefont {C.~D.~P.}\ \bibnamefont
				{Levy}}, \bibinfo {author} {\bibfnamefont {M.~R.}\ \bibnamefont {Pearson}},
			\bibinfo {author} {\bibfnamefont {T.}~\bibnamefont {Buck}}, \bibinfo {author}
			{\bibfnamefont {K.~H.}\ \bibnamefont {Chow}}, \bibinfo {author}
			{\bibfnamefont {A.~N.}\ \bibnamefont {Hariwal}}, \bibinfo {author}
			{\bibfnamefont {R.~F.}\ \bibnamefont {Kiefl}}, \bibinfo {author}
			{\bibfnamefont {F.~H.}\ \bibnamefont {McGee}}, \bibinfo {author}
			{\bibfnamefont {G.~D.}\ \bibnamefont {Morris}},\ and\ \bibinfo {author}
			{\bibfnamefont {D.}~\bibnamefont {Wang}},\ }\bibfield  {title} {\enquote
			{\bibinfo {title} {The {I}nitial {S}tate of {O}ptically {P}olarized
					\ch{^{8}Li^{+}} from the $\beta$-{NMR} in {B}ismuth},}\ }\href
		{https://doi.org/10.1088/1742-6596/551/1/012059} {\bibfield  {journal}
			{\bibinfo  {journal} {J. Phys. Conf. Ser.}\ }\textbf {\bibinfo {volume}
				{551}},\ \bibinfo {pages} {012059} (\bibinfo {year} {2014})}\BibitemShut
		{NoStop}%
		\bibitem [{GPT()}]{GPT}%
		\BibitemOpen
		GPT,\ \href {http://www.pulsar.nl/gpt} {\enquote {\bibinfo {title} {General
					particle tracer},}\ }\BibitemShut {NoStop}%
		\bibitem [{\citenamefont {Heighway}\ and\ \citenamefont
			{Hutcheon}(1981)}]{Heighway1981}%
		\BibitemOpen
		\bibfield  {author} {\bibinfo {author} {\bibfnamefont {E.}~\bibnamefont
				{Heighway}}\ and\ \bibinfo {author} {\bibfnamefont {R.}~\bibnamefont
				{Hutcheon}},\ }\bibfield  {title} {\enquote {\bibinfo {title} {Transoptr —
					a second order beam transport design code with optimization and
					constraints},}\ }\href
		{https://doi.org/https://doi.org/10.1016/0029-554X(81)90474-2} {\bibfield
			{journal} {\bibinfo  {journal} {Nucl. Instrum. Methods Phys. Res.}\ }\textbf
			{\bibinfo {volume} {187}},\ \bibinfo {pages} {89--95} (\bibinfo {year}
			{1981})}\BibitemShut {NoStop}%
		\bibitem [{\citenamefont {Baartman}(2022)}]{TRANSOPTR_Manual}%
		\BibitemOpen
		\bibfield  {author} {\bibinfo {author} {\bibfnamefont {R.}~\bibnamefont
				{Baartman}},\ }\href
		{http://lin12.triumf.ca/text/design_notes/transoptr-manual.pdf} {\emph
			{\bibinfo {title} {TRANSOPTR Reference Manual}}},\ \bibinfo {organization}
		{TRIUMF} (\bibinfo {year} {2022})\BibitemShut {NoStop}%
		\bibitem [{\citenamefont {Dalesio}\ \emph {et~al.}(1994)\citenamefont
			{Dalesio}, \citenamefont {Hill}, \citenamefont {Kraimer}, \citenamefont
			{Lewis}, \citenamefont {Murray}, \citenamefont {Hunt}, \citenamefont
			{Watson}, \citenamefont {Clausen},\ and\ \citenamefont
			{Dalesio}}]{Dalesio1994}%
		\BibitemOpen
		\bibfield  {author} {\bibinfo {author} {\bibfnamefont {L.~R.}\ \bibnamefont
				{Dalesio}}, \bibinfo {author} {\bibfnamefont {J.~O.}\ \bibnamefont {Hill}},
			\bibinfo {author} {\bibfnamefont {M.}~\bibnamefont {Kraimer}}, \bibinfo
			{author} {\bibfnamefont {S.}~\bibnamefont {Lewis}}, \bibinfo {author}
			{\bibfnamefont {D.}~\bibnamefont {Murray}}, \bibinfo {author} {\bibfnamefont
				{S.}~\bibnamefont {Hunt}}, \bibinfo {author} {\bibfnamefont {W.}~\bibnamefont
				{Watson}}, \bibinfo {author} {\bibfnamefont {M.}~\bibnamefont {Clausen}},\
			and\ \bibinfo {author} {\bibfnamefont {J.}~\bibnamefont {Dalesio}},\
		}\bibfield  {title} {\enquote {\bibinfo {title} {The experimental physics and
					industrial control system architecture: past, present, and future},}\ }\href
		{https://doi.org/https://doi.org/10.1016/0168-9002(94)91493-1} {\bibfield
			{journal} {\bibinfo  {journal} {Nucl. Instrum. Methods Phys. Res., Sect. A}\
			}\textbf {\bibinfo {volume} {352}},\ \bibinfo {pages} {179--184} (\bibinfo
			{year} {1994})}\BibitemShut {NoStop}%
		\bibitem [{EPICS()}]{EPICS}%
		\BibitemOpen
		EPICS,\ \href@noop {} {\enquote {\bibinfo {title} {Experimental physics and
					industrial control system},}\ }\bibinfo {note}
		{\url{http://www.aps.anl.gov/epics/}}\BibitemShut {NoStop}%
		\bibitem [{\citenamefont {Salman}\ \emph {et~al.}(2014)\citenamefont {Salman},
			\citenamefont {Chow}, \citenamefont {Hossain}, \citenamefont {Kiefl},
			\citenamefont {Levy}, \citenamefont {Parolin}, \citenamefont {Pearson},
			\citenamefont {Saadaoui}, \citenamefont {Wang},\ and\ \citenamefont
			{MacFarlane}}]{Salman2014_sapphire}%
		\BibitemOpen
		\bibfield  {author} {\bibinfo {author} {\bibfnamefont {Z.}~\bibnamefont
				{Salman}}, \bibinfo {author} {\bibfnamefont {K.~H.}\ \bibnamefont {Chow}},
			\bibinfo {author} {\bibfnamefont {M.~D.}\ \bibnamefont {Hossain}}, \bibinfo
			{author} {\bibfnamefont {R.~F.}\ \bibnamefont {Kiefl}}, \bibinfo {author}
			{\bibfnamefont {C.~D.~P.}\ \bibnamefont {Levy}}, \bibinfo {author}
			{\bibfnamefont {T.~J.}\ \bibnamefont {Parolin}}, \bibinfo {author}
			{\bibfnamefont {M.~R.}\ \bibnamefont {Pearson}}, \bibinfo {author}
			{\bibfnamefont {H.}~\bibnamefont {Saadaoui}}, \bibinfo {author}
			{\bibfnamefont {D.}~\bibnamefont {Wang}},\ and\ \bibinfo {author}
			{\bibfnamefont {W.~A.}\ \bibnamefont {MacFarlane}},\ }\bibfield  {title}
		{\enquote {\bibinfo {title} {$\beta$-detected nuclear quadrupole resonance
					and relaxation of $^8${Li}$^+$ in sapphire},}\ }\href
		{https://doi.org/10.1088/1742-6596/551/1/012034} {\bibfield  {journal}
			{\bibinfo  {journal} {J. Phys. Conf. Ser.}\ }\textbf {\bibinfo {volume}
				{551}},\ \bibinfo {pages} {012034} (\bibinfo {year} {2014})}\BibitemShut
		{NoStop}%
		\bibitem [{OPERA()}]{OPERA}%
		\BibitemOpen
		OPERA,\ \href {https://www.3ds.com/products-services/simulia/products/opera/}
		{\enquote {\bibinfo {title} {Opera-3d},}\ }\BibitemShut {NoStop}%
		\bibitem [{CST()}]{CST}%
		\BibitemOpen
		CST,\ \href
		{https://www.3ds.com/products-services/simulia/products/cst-studio-suite/}
		{\enquote {\bibinfo {title} {{C}{S}{T} {S}tudio},}\ }\BibitemShut {NoStop}%
		\bibitem [{\citenamefont {Stone}(2019)}]{IAEA_TableNuclearMom_gyro_Stone2019}%
		\BibitemOpen
		\bibfield  {author} {\bibinfo {author} {\bibfnamefont {N.~J.}\ \bibnamefont
				{Stone}},\ }\href
		{http://inis.iaea.org/search/search.aspx?orig_q=RN:51052833} {\enquote
			{\bibinfo {title} {Table of recommended nuclear magnetic dipole moments},}\
		}\bibinfo {type} {Tech. Rep.}\ \bibinfo {number} {INDC(NDS)--0794}\ (\bibinfo
		{address} {International Atomic Energy Agency (IAEA)},\ \bibinfo {year}
		{2019})\BibitemShut {NoStop}%
		\bibitem [{\citenamefont {Slichter}(1990)}]{SlichterBook}%
		\BibitemOpen
		\bibfield  {author} {\bibinfo {author} {\bibfnamefont {C.~P.}\ \bibnamefont
				{Slichter}},\ }\enquote {\bibinfo {title} {Spin-lattice relaxation and
				motional narrowing of resonance lines},}\ in\ \href
		{https://doi.org/10.1007/978-3-662-09441-9_5} {\emph {\bibinfo {booktitle}
				{Principles of Magnetic Resonance}}}\ (\bibinfo  {publisher} {Springer Berlin
			Heidelberg},\ \bibinfo {address} {Berlin, Heidelberg},\ \bibinfo {year}
		{1990})\ pp.\ \bibinfo {pages} {145--218}\BibitemShut {NoStop}%
		\bibitem [{\citenamefont {Fujimoto}(2020)}]{bfit_arXiV}%
		\BibitemOpen
		\bibfield  {author} {\bibinfo {author} {\bibfnamefont {D.}~\bibnamefont
				{Fujimoto}},\ }\href {https://doi.org/10.48550/ARXIV.2004.10395} {\enquote
			{\bibinfo {title} {Digging into mud with python: mudpy, bdata, and bfit},}\ }
		(\bibinfo {year} {2020}),\ \Eprint {https://arxiv.org/abs/2004.10395}
		{arXiv:2004.10395} \BibitemShut {NoStop}%
		\bibitem [{\citenamefont {Fujimoto}(2021)}]{bfit_paper}%
		\BibitemOpen
		\bibfield  {author} {\bibinfo {author} {\bibfnamefont {D.}~\bibnamefont
				{Fujimoto}},\ }\bibfield  {title} {\enquote {\bibinfo {title} {bfit: A python
					application for beta-detected nmr},}\ }\href
		{https://doi.org/10.21105/joss.03598} {\bibfield  {journal} {\bibinfo
				{journal} {Journal of Open Source Software}\ }\textbf {\bibinfo {volume}
				{6}},\ \bibinfo {pages} {3598} (\bibinfo {year} {2021})}\BibitemShut
		{NoStop}%
		\bibitem [{\citenamefont {MacFarlane}\ \emph {et~al.}(2018)\citenamefont
			{MacFarlane}, \citenamefont {Chow}, \citenamefont {Hossain}, \citenamefont
			{Karner}, \citenamefont {Kiefl}, \citenamefont {McFadden}, \citenamefont
			{Morris}, \citenamefont {Saadaoui},\ and\ \citenamefont
			{Salman}}]{MacFarlane_Gold}%
		\BibitemOpen
		\bibfield  {author} {\bibinfo {author} {\bibfnamefont {W.~A.}\ \bibnamefont
				{MacFarlane}}, \bibinfo {author} {\bibfnamefont {K.~H.}\ \bibnamefont
				{Chow}}, \bibinfo {author} {\bibfnamefont {M.~D.}\ \bibnamefont {Hossain}},
			\bibinfo {author} {\bibfnamefont {V.~L.}\ \bibnamefont {Karner}}, \bibinfo
			{author} {\bibfnamefont {R.~F.}\ \bibnamefont {Kiefl}}, \bibinfo {author}
			{\bibfnamefont {R.~M.~L.}\ \bibnamefont {McFadden}}, \bibinfo {author}
			{\bibfnamefont {G.~D.}\ \bibnamefont {Morris}}, \bibinfo {author}
			{\bibfnamefont {H.}~\bibnamefont {Saadaoui}},\ and\ \bibinfo {author}
			{\bibfnamefont {Z.}~\bibnamefont {Salman}},\ }\bibfield  {title} {\enquote
			{\bibinfo {title} {The spin relaxation of \ch{^{8}Li^{+}} in gold at low
					magnetic field},}\ }\href {https://doi.org/10.7566/JPSCP.21.011020}
		{\bibfield  {journal} {\bibinfo  {journal} {JPS Conf. Proc.}\ }\textbf
			{\bibinfo {volume} {21}},\ \bibinfo {pages} {011020} (\bibinfo {year}
			{2018})}\BibitemShut {NoStop}%
		\bibitem [{\citenamefont {Parolin}\ \emph {et~al.}(2008)\citenamefont
			{Parolin}, \citenamefont {Salman}, \citenamefont {Chow}, \citenamefont
			{Song}, \citenamefont {Valiani}, \citenamefont {Saadaoui}, \citenamefont
			{O'Halloran}, \citenamefont {Hossain}, \citenamefont {Keeler}, \citenamefont
			{Kiefl}, \citenamefont {Kreitzman}, \citenamefont {Levy}, \citenamefont
			{Miller}, \citenamefont {Morris}, \citenamefont {Pearson}, \citenamefont
			{Smadella}, \citenamefont {Wang}, \citenamefont {Xu},\ and\ \citenamefont
			{MacFarlane}}]{Parolin_Gold}%
		\BibitemOpen
		\bibfield  {author} {\bibinfo {author} {\bibfnamefont {T.~J.}\ \bibnamefont
				{Parolin}}, \bibinfo {author} {\bibfnamefont {Z.}~\bibnamefont {Salman}},
			\bibinfo {author} {\bibfnamefont {K.~H.}\ \bibnamefont {Chow}}, \bibinfo
			{author} {\bibfnamefont {Q.}~\bibnamefont {Song}}, \bibinfo {author}
			{\bibfnamefont {J.}~\bibnamefont {Valiani}}, \bibinfo {author} {\bibfnamefont
				{H.}~\bibnamefont {Saadaoui}}, \bibinfo {author} {\bibfnamefont
				{A.}~\bibnamefont {O'Halloran}}, \bibinfo {author} {\bibfnamefont {M.~D.}\
				\bibnamefont {Hossain}}, \bibinfo {author} {\bibfnamefont {T.~A.}\
				\bibnamefont {Keeler}}, \bibinfo {author} {\bibfnamefont {R.~F.}\
				\bibnamefont {Kiefl}}, \bibinfo {author} {\bibfnamefont {S.~R.}\ \bibnamefont
				{Kreitzman}}, \bibinfo {author} {\bibfnamefont {C.~D.~P.}\ \bibnamefont
				{Levy}}, \bibinfo {author} {\bibfnamefont {R.~I.}\ \bibnamefont {Miller}},
			\bibinfo {author} {\bibfnamefont {G.~D.}\ \bibnamefont {Morris}}, \bibinfo
			{author} {\bibfnamefont {M.~R.}\ \bibnamefont {Pearson}}, \bibinfo {author}
			{\bibfnamefont {M.}~\bibnamefont {Smadella}}, \bibinfo {author}
			{\bibfnamefont {D.}~\bibnamefont {Wang}}, \bibinfo {author} {\bibfnamefont
				{M.}~\bibnamefont {Xu}},\ and\ \bibinfo {author} {\bibfnamefont {W.~A.}\
				\bibnamefont {MacFarlane}},\ }\bibfield  {title} {\enquote {\bibinfo {title}
				{High resolution $\ensuremath{\beta}\text{-NMR}$ study of
					${^{8}\text{L}\text{i}}^{+}$ implanted in gold},}\ }\href
		{https://doi.org/10.1103/PhysRevB.77.214107} {\bibfield  {journal} {\bibinfo
				{journal} {Phys. Rev. B}\ }\textbf {\bibinfo {volume} {77}},\ \bibinfo
			{pages} {214107} (\bibinfo {year} {2008})}\BibitemShut {NoStop}%
		\bibitem [{\citenamefont {Parolin}\ \emph {et~al.}(2009)\citenamefont
			{Parolin}, \citenamefont {Shi}, \citenamefont {Salman}, \citenamefont {Chow},
			\citenamefont {Dosanjh}, \citenamefont {Saadaoui}, \citenamefont {Song},
			\citenamefont {Hossain}, \citenamefont {Kiefl}, \citenamefont {Levy},
			\citenamefont {Pearson},\ and\ \citenamefont {MacFarlane}}]{Parolin_Nb}%
		\BibitemOpen
		\bibfield  {author} {\bibinfo {author} {\bibfnamefont {T.~J.}\ \bibnamefont
				{Parolin}}, \bibinfo {author} {\bibfnamefont {J.}~\bibnamefont {Shi}},
			\bibinfo {author} {\bibfnamefont {Z.}~\bibnamefont {Salman}}, \bibinfo
			{author} {\bibfnamefont {K.~H.}\ \bibnamefont {Chow}}, \bibinfo {author}
			{\bibfnamefont {P.}~\bibnamefont {Dosanjh}}, \bibinfo {author} {\bibfnamefont
				{H.}~\bibnamefont {Saadaoui}}, \bibinfo {author} {\bibfnamefont
				{Q.}~\bibnamefont {Song}}, \bibinfo {author} {\bibfnamefont {M.~D.}\
				\bibnamefont {Hossain}}, \bibinfo {author} {\bibfnamefont {R.~F.}\
				\bibnamefont {Kiefl}}, \bibinfo {author} {\bibfnamefont {C.~D.~P.}\
				\bibnamefont {Levy}}, \bibinfo {author} {\bibfnamefont {M.~R.}\ \bibnamefont
				{Pearson}},\ and\ \bibinfo {author} {\bibfnamefont {W.~A.}\ \bibnamefont
				{MacFarlane}},\ }\bibfield  {title} {\enquote {\bibinfo {title} {Nuclear
					magnetic resonance study of li implanted in a thin film of niobium},}\ }\href
		{https://doi.org/10.1103/PhysRevB.80.174109} {\bibfield  {journal} {\bibinfo
				{journal} {Phys. Rev. B}\ }\textbf {\bibinfo {volume} {80}},\ \bibinfo
			{pages} {174109} (\bibinfo {year} {2009})}\BibitemShut {NoStop}%
		\bibitem [{\citenamefont {{Ziegler}}, \citenamefont {{Ziegler}},\ and\
			\citenamefont {{Biersack}}(2010)}]{2010SRIM}%
		\BibitemOpen
		\bibfield  {author} {\bibinfo {author} {\bibfnamefont {J.~F.}\ \bibnamefont
				{{Ziegler}}}, \bibinfo {author} {\bibfnamefont {M.~D.}\ \bibnamefont
				{{Ziegler}}},\ and\ \bibinfo {author} {\bibfnamefont {J.~P.}\ \bibnamefont
				{{Biersack}}},\ }\bibfield  {title} {\enquote {\bibinfo {title} {{SRIM - The
						stopping and range of ions in matter (2010)}},}\ }\href
		{https://doi.org/10.1016/j.nimb.2010.02.091} {\bibfield  {journal} {\bibinfo
				{journal} {Nucl. Instrum. Methods Phys. Res., Sect. B}\ }\textbf {\bibinfo
				{volume} {268}},\ \bibinfo {pages} {1818--1823} (\bibinfo {year}
			{2010})}\BibitemShut {NoStop}%
		\bibitem [{\citenamefont {Hossain}\ \emph {et~al.}(2009)\citenamefont
			{Hossain}, \citenamefont {Salman}, \citenamefont {Wang}, \citenamefont
			{Chow}, \citenamefont {Kreitzman}, \citenamefont {Keeler}, \citenamefont
			{Levy}, \citenamefont {MacFarlane}, \citenamefont {Miller}, \citenamefont
			{Morris}, \citenamefont {Parolin}, \citenamefont {Pearson}, \citenamefont
			{Saadaoui},\ and\ \citenamefont {Kiefl}}]{Hossain2009}%
		\BibitemOpen
		\bibfield  {author} {\bibinfo {author} {\bibfnamefont {M.~D.}\ \bibnamefont
				{Hossain}}, \bibinfo {author} {\bibfnamefont {Z.}~\bibnamefont {Salman}},
			\bibinfo {author} {\bibfnamefont {D.}~\bibnamefont {Wang}}, \bibinfo {author}
			{\bibfnamefont {K.~H.}\ \bibnamefont {Chow}}, \bibinfo {author}
			{\bibfnamefont {S.}~\bibnamefont {Kreitzman}}, \bibinfo {author}
			{\bibfnamefont {T.~A.}\ \bibnamefont {Keeler}}, \bibinfo {author}
			{\bibfnamefont {C.~D.~P.}\ \bibnamefont {Levy}}, \bibinfo {author}
			{\bibfnamefont {W.~A.}\ \bibnamefont {MacFarlane}}, \bibinfo {author}
			{\bibfnamefont {R.~I.}\ \bibnamefont {Miller}}, \bibinfo {author}
			{\bibfnamefont {G.~D.}\ \bibnamefont {Morris}}, \bibinfo {author}
			{\bibfnamefont {T.~J.}\ \bibnamefont {Parolin}}, \bibinfo {author}
			{\bibfnamefont {M.}~\bibnamefont {Pearson}}, \bibinfo {author} {\bibfnamefont
				{H.}~\bibnamefont {Saadaoui}},\ and\ \bibinfo {author} {\bibfnamefont
				{R.~F.}\ \bibnamefont {Kiefl}},\ }\bibfield  {title} {\enquote {\bibinfo
				{title} {Low-field cross spin relaxation of $^{8}\text{L}\text{i}$ in
					superconducting \ch{NbSe2}},}\ }\href
		{https://doi.org/10.1103/PhysRevB.79.144518} {\bibfield  {journal} {\bibinfo
				{journal} {Phys. Rev. B}\ }\textbf {\bibinfo {volume} {79}},\ \bibinfo
			{pages} {144518} (\bibinfo {year} {2009})}\BibitemShut {NoStop}%
		\bibitem [{\citenamefont {Bloembergen}, \citenamefont {Purcell},\ and\
			\citenamefont {Pound}(1948)}]{BPP_orig}%
		\BibitemOpen
		\bibfield  {author} {\bibinfo {author} {\bibfnamefont {N.}~\bibnamefont
				{Bloembergen}}, \bibinfo {author} {\bibfnamefont {E.~M.}\ \bibnamefont
				{Purcell}},\ and\ \bibinfo {author} {\bibfnamefont {R.~V.}\ \bibnamefont
				{Pound}},\ }\bibfield  {title} {\enquote {\bibinfo {title} {Relaxation
					effects in nuclear magnetic resonance absorption},}\ }\href
		{https://doi.org/10.1103/PhysRev.73.679} {\bibfield  {journal} {\bibinfo
				{journal} {Phys. Rev.}\ }\textbf {\bibinfo {volume} {73}},\ \bibinfo {pages}
			{679--712} (\bibinfo {year} {1948})}\BibitemShut {NoStop}%
		\bibitem [{\citenamefont {McKenzie}\ \emph {et~al.}(2014)\citenamefont
			{McKenzie}, \citenamefont {Harada}, \citenamefont {Kiefl}, \citenamefont
			{Levy}, \citenamefont {MacFarlane}, \citenamefont {Morris}, \citenamefont
			{Ogata}, \citenamefont {Pearson},\ and\ \citenamefont
			{Sugiyama}}]{2014-McKenzie-JACS-136-7833}%
		\BibitemOpen
		\bibfield  {author} {\bibinfo {author} {\bibfnamefont {I.}~\bibnamefont
				{McKenzie}}, \bibinfo {author} {\bibfnamefont {M.}~\bibnamefont {Harada}},
			\bibinfo {author} {\bibfnamefont {R.~F.}\ \bibnamefont {Kiefl}}, \bibinfo
			{author} {\bibfnamefont {C.~D.~P.}\ \bibnamefont {Levy}}, \bibinfo {author}
			{\bibfnamefont {W.~A.}\ \bibnamefont {MacFarlane}}, \bibinfo {author}
			{\bibfnamefont {G.~D.}\ \bibnamefont {Morris}}, \bibinfo {author}
			{\bibfnamefont {S.-I.}\ \bibnamefont {Ogata}}, \bibinfo {author}
			{\bibfnamefont {M.~R.}\ \bibnamefont {Pearson}},\ and\ \bibinfo {author}
			{\bibfnamefont {J.}~\bibnamefont {Sugiyama}},\ }\bibfield  {title} {\enquote
			{\bibinfo {title} {{{$\beta$}-NMR} measurements of lithium ion transport in
					thin films of pure and lithium-salt-doped poly(ethylene oxide)},}\ }\href
		{https://doi.org/10.1021/ja503066a} {\bibfield  {journal} {\bibinfo
				{journal} {J. Am. Chem. Soc.}\ }\textbf {\bibinfo {volume} {136}},\ \bibinfo
			{pages} {7833--7836} (\bibinfo {year} {2014})}\BibitemShut {NoStop}%
		\bibitem [{\citenamefont {McKenzie}\ \emph {et~al.}(2017)\citenamefont
			{McKenzie}, \citenamefont {Cortie}, \citenamefont {Harada}, \citenamefont
			{Kiefl}, \citenamefont {Levy}, \citenamefont {MacFarlane}, \citenamefont
			{McFadden}, \citenamefont {Morris}, \citenamefont {Ogata}, \citenamefont
			{Pearson},\ and\ \citenamefont {Sugiyama}}]{2017-McKenzie-JCP-146-244903}%
		\BibitemOpen
		\bibfield  {author} {\bibinfo {author} {\bibfnamefont {I.}~\bibnamefont
				{McKenzie}}, \bibinfo {author} {\bibfnamefont {D.~L.}\ \bibnamefont
				{Cortie}}, \bibinfo {author} {\bibfnamefont {M.}~\bibnamefont {Harada}},
			\bibinfo {author} {\bibfnamefont {R.~F.}\ \bibnamefont {Kiefl}}, \bibinfo
			{author} {\bibfnamefont {C.~D.~P.}\ \bibnamefont {Levy}}, \bibinfo {author}
			{\bibfnamefont {W.~A.}\ \bibnamefont {MacFarlane}}, \bibinfo {author}
			{\bibfnamefont {R.~M.~L.}\ \bibnamefont {McFadden}}, \bibinfo {author}
			{\bibfnamefont {G.~D.}\ \bibnamefont {Morris}}, \bibinfo {author}
			{\bibfnamefont {S.-I.}\ \bibnamefont {Ogata}}, \bibinfo {author}
			{\bibfnamefont {M.~R.}\ \bibnamefont {Pearson}},\ and\ \bibinfo {author}
			{\bibfnamefont {J.}~\bibnamefont {Sugiyama}},\ }\bibfield  {title} {\enquote
			{\bibinfo {title} {{{$\beta$}-NMR} measurements of molecular-scale
					lithium-ion dynamics in poly(ethylene oxide)-lithium-salt thin films},}\
		}\href {https://doi.org/10.1063/1.4989866} {\bibfield  {journal} {\bibinfo
				{journal} {J. Chem. Phys.}\ }\textbf {\bibinfo {volume} {146}},\ \bibinfo
			{pages} {244903} (\bibinfo {year} {2017})}\BibitemShut {NoStop}%
		\bibitem [{\citenamefont {McFadden}\ \emph {et~al.}(2017)\citenamefont
			{McFadden}, \citenamefont {Buck}, \citenamefont {Chatzichristos},
			\citenamefont {Chen}, \citenamefont {Chow}, \citenamefont {Cortie},
			\citenamefont {Dehn}, \citenamefont {Karner}, \citenamefont {Koumoulis},
			\citenamefont {Levy}, \citenamefont {Li}, \citenamefont {McKenzie},
			\citenamefont {Merkle}, \citenamefont {Morris}, \citenamefont {Pearson},
			\citenamefont {Salman}, \citenamefont {Samuelis}, \citenamefont {Stachura},
			\citenamefont {Xiao}, \citenamefont {Maier}, \citenamefont {Kiefl},\ and\
			\citenamefont {MacFarlane}}]{2017-McFadden-CM-29-10187}%
		\BibitemOpen
		\bibfield  {author} {\bibinfo {author} {\bibfnamefont {R.~M.~L.}\
				\bibnamefont {McFadden}}, \bibinfo {author} {\bibfnamefont {T.~J.}\
				\bibnamefont {Buck}}, \bibinfo {author} {\bibfnamefont {A.}~\bibnamefont
				{Chatzichristos}}, \bibinfo {author} {\bibfnamefont {C.-C.}\ \bibnamefont
				{Chen}}, \bibinfo {author} {\bibfnamefont {K.~H.}\ \bibnamefont {Chow}},
			\bibinfo {author} {\bibfnamefont {D.~L.}\ \bibnamefont {Cortie}}, \bibinfo
			{author} {\bibfnamefont {M.~H.}\ \bibnamefont {Dehn}}, \bibinfo {author}
			{\bibfnamefont {V.~L.}\ \bibnamefont {Karner}}, \bibinfo {author}
			{\bibfnamefont {D.}~\bibnamefont {Koumoulis}}, \bibinfo {author}
			{\bibfnamefont {C.~D.~P.}\ \bibnamefont {Levy}}, \bibinfo {author}
			{\bibfnamefont {C.}~\bibnamefont {Li}}, \bibinfo {author} {\bibfnamefont
				{I.}~\bibnamefont {McKenzie}}, \bibinfo {author} {\bibfnamefont
				{R.}~\bibnamefont {Merkle}}, \bibinfo {author} {\bibfnamefont {G.~D.}\
				\bibnamefont {Morris}}, \bibinfo {author} {\bibfnamefont {M.~R.}\
				\bibnamefont {Pearson}}, \bibinfo {author} {\bibfnamefont {Z.}~\bibnamefont
				{Salman}}, \bibinfo {author} {\bibfnamefont {D.}~\bibnamefont {Samuelis}},
			\bibinfo {author} {\bibfnamefont {M.}~\bibnamefont {Stachura}}, \bibinfo
			{author} {\bibfnamefont {J.}~\bibnamefont {Xiao}}, \bibinfo {author}
			{\bibfnamefont {J.}~\bibnamefont {Maier}}, \bibinfo {author} {\bibfnamefont
				{R.~F.}\ \bibnamefont {Kiefl}},\ and\ \bibinfo {author} {\bibfnamefont
				{W.~A.}\ \bibnamefont {MacFarlane}},\ }\bibfield  {title} {\enquote {\bibinfo
				{title} {Microscopic dynamics of \ch{Li^{+}} in rutile \ch{TiO2} revealed by
					\ch{^{8}Li} {$\beta$}-detected nuclear magnetic resonance},}\ }\href
		{https://doi.org/10.1021/acs.chemmater.7b04093} {\bibfield  {journal}
			{\bibinfo  {journal} {Chem. Mater.}\ }\textbf {\bibinfo {volume} {29}},\
			\bibinfo {pages} {10187--10197} (\bibinfo {year} {2017})}\BibitemShut
		{NoStop}%
		\bibitem [{\citenamefont {McFadden}\ \emph {et~al.}(2019)\citenamefont
			{McFadden}, \citenamefont {Chatzichristos}, \citenamefont {Chow},
			\citenamefont {Cortie}, \citenamefont {Dehn}, \citenamefont {Fujimoto},
			\citenamefont {Hossain}, \citenamefont {Ji}, \citenamefont {Karner},
			\citenamefont {Kiefl}, \citenamefont {Levy}, \citenamefont {Li},
			\citenamefont {McKenzie}, \citenamefont {Morris}, \citenamefont {Ofer},
			\citenamefont {Pearson}, \citenamefont {Stachura}, \citenamefont {Cava},\
			and\ \citenamefont {MacFarlane}}]{2019-McFadden-PRB-99-125201}%
		\BibitemOpen
		\bibfield  {author} {\bibinfo {author} {\bibfnamefont {R.~M.~L.}\
				\bibnamefont {McFadden}}, \bibinfo {author} {\bibfnamefont {A.}~\bibnamefont
				{Chatzichristos}}, \bibinfo {author} {\bibfnamefont {K.~H.}\ \bibnamefont
				{Chow}}, \bibinfo {author} {\bibfnamefont {D.~L.}\ \bibnamefont {Cortie}},
			\bibinfo {author} {\bibfnamefont {M.~H.}\ \bibnamefont {Dehn}}, \bibinfo
			{author} {\bibfnamefont {D.}~\bibnamefont {Fujimoto}}, \bibinfo {author}
			{\bibfnamefont {M.~D.}\ \bibnamefont {Hossain}}, \bibinfo {author}
			{\bibfnamefont {H.}~\bibnamefont {Ji}}, \bibinfo {author} {\bibfnamefont
				{V.~L.}\ \bibnamefont {Karner}}, \bibinfo {author} {\bibfnamefont {R.~F.}\
				\bibnamefont {Kiefl}}, \bibinfo {author} {\bibfnamefont {C.~D.~P.}\
				\bibnamefont {Levy}}, \bibinfo {author} {\bibfnamefont {R.}~\bibnamefont
				{Li}}, \bibinfo {author} {\bibfnamefont {I.}~\bibnamefont {McKenzie}},
			\bibinfo {author} {\bibfnamefont {G.~D.}\ \bibnamefont {Morris}}, \bibinfo
			{author} {\bibfnamefont {O.}~\bibnamefont {Ofer}}, \bibinfo {author}
			{\bibfnamefont {M.~R.}\ \bibnamefont {Pearson}}, \bibinfo {author}
			{\bibfnamefont {M.}~\bibnamefont {Stachura}}, \bibinfo {author}
			{\bibfnamefont {R.~J.}\ \bibnamefont {Cava}},\ and\ \bibinfo {author}
			{\bibfnamefont {W.~A.}\ \bibnamefont {MacFarlane}},\ }\bibfield  {title}
		{\enquote {\bibinfo {title} {Ionic and electronic properties of the
					topological insulator \ch{Bi2Te2Se} investigated via {$\beta$}-detected
					nuclear magnetic relaxation and resonance of \ch{^{8}Li}},}\ }\href
		{https://doi.org/10.1103/PhysRevB.99.125201} {\bibfield  {journal} {\bibinfo
				{journal} {Phys. Rev. B}\ }\textbf {\bibinfo {volume} {99}},\ \bibinfo
			{pages} {125201} (\bibinfo {year} {2019})}\BibitemShut {NoStop}%
		\bibitem [{\citenamefont {McFadden}\ \emph {et~al.}(2020)\citenamefont
			{McFadden}, \citenamefont {Chatzichristos}, \citenamefont {Cortie},
			\citenamefont {Fujimoto}, \citenamefont {Hor}, \citenamefont {Ji},
			\citenamefont {Karner}, \citenamefont {Kiefl}, \citenamefont {Levy},
			\citenamefont {Li}, \citenamefont {McKenzie}, \citenamefont {Morris},
			\citenamefont {Pearson}, \citenamefont {Stachura}, \citenamefont {Cava},\
			and\ \citenamefont {MacFarlane}}]{2020-McFadden-PRB-102-235206}%
		\BibitemOpen
		\bibfield  {author} {\bibinfo {author} {\bibfnamefont {R.~M.~L.}\
				\bibnamefont {McFadden}}, \bibinfo {author} {\bibfnamefont {A.}~\bibnamefont
				{Chatzichristos}}, \bibinfo {author} {\bibfnamefont {D.~L.}\ \bibnamefont
				{Cortie}}, \bibinfo {author} {\bibfnamefont {D.}~\bibnamefont {Fujimoto}},
			\bibinfo {author} {\bibfnamefont {Y.~S.}\ \bibnamefont {Hor}}, \bibinfo
			{author} {\bibfnamefont {H.}~\bibnamefont {Ji}}, \bibinfo {author}
			{\bibfnamefont {V.~L.}\ \bibnamefont {Karner}}, \bibinfo {author}
			{\bibfnamefont {R.~F.}\ \bibnamefont {Kiefl}}, \bibinfo {author}
			{\bibfnamefont {C.~D.~P.}\ \bibnamefont {Levy}}, \bibinfo {author}
			{\bibfnamefont {R.}~\bibnamefont {Li}}, \bibinfo {author} {\bibfnamefont
				{I.}~\bibnamefont {McKenzie}}, \bibinfo {author} {\bibfnamefont {G.~D.}\
				\bibnamefont {Morris}}, \bibinfo {author} {\bibfnamefont {M.~R.}\
				\bibnamefont {Pearson}}, \bibinfo {author} {\bibfnamefont {M.}~\bibnamefont
				{Stachura}}, \bibinfo {author} {\bibfnamefont {R.~J.}\ \bibnamefont {Cava}},\
			and\ \bibinfo {author} {\bibfnamefont {W.~A.}\ \bibnamefont {MacFarlane}},\
		}\bibfield  {title} {\enquote {\bibinfo {title} {Local electronic and
					magnetic properties of the doped topological insulators \ch{Bi2Se3:Ca} and
					\ch{Bi2Te3:Mn} investigated using ion-implanted \ch{^8Li} {{$\beta$}-NMR}},}\
		}\href {https://doi.org/10.1103/PhysRevB.102.235206} {\bibfield  {journal}
			{\bibinfo  {journal} {Phys. Rev. B}\ }\textbf {\bibinfo {volume} {102}},\
			\bibinfo {pages} {235206} (\bibinfo {year} {2020})}\BibitemShut {NoStop}%
		\bibitem [{\citenamefont {McGee}\ \emph {et~al.}(2014)\citenamefont {McGee},
			\citenamefont {McKenzie}, \citenamefont {Buck}, \citenamefont {Daley},
			\citenamefont {Forrest}, \citenamefont {Harada}, \citenamefont {Kiefl},
			\citenamefont {Levy}, \citenamefont {Morris}, \citenamefont {Pearson},
			\citenamefont {Sugiyama}, \citenamefont {Wang},\ and\ \citenamefont
			{MacFarlane}}]{2014-McGee-JPCS-551-012039}%
		\BibitemOpen
		\bibfield  {author} {\bibinfo {author} {\bibfnamefont {F.~H.}\ \bibnamefont
				{McGee}}, \bibinfo {author} {\bibfnamefont {I.}~\bibnamefont {McKenzie}},
			\bibinfo {author} {\bibfnamefont {T.}~\bibnamefont {Buck}}, \bibinfo {author}
			{\bibfnamefont {C.~R.}\ \bibnamefont {Daley}}, \bibinfo {author}
			{\bibfnamefont {J.~A.}\ \bibnamefont {Forrest}}, \bibinfo {author}
			{\bibfnamefont {M.}~\bibnamefont {Harada}}, \bibinfo {author} {\bibfnamefont
				{R.~F.}\ \bibnamefont {Kiefl}}, \bibinfo {author} {\bibfnamefont {C.~D.~P.}\
				\bibnamefont {Levy}}, \bibinfo {author} {\bibfnamefont {G.~D.}\ \bibnamefont
				{Morris}}, \bibinfo {author} {\bibfnamefont {M.~R.}\ \bibnamefont {Pearson}},
			\bibinfo {author} {\bibfnamefont {J.}~\bibnamefont {Sugiyama}}, \bibinfo
			{author} {\bibfnamefont {D.}~\bibnamefont {Wang}},\ and\ \bibinfo {author}
			{\bibfnamefont {W.~A.}\ \bibnamefont {MacFarlane}},\ }\bibfield  {title}
		{\enquote {\bibinfo {title} {A brief survey of {$\beta$}-detected {NMR} of
					implanted \ch{^{8}Li^{+}} in organic polymers},}\ }\href
		{https://doi.org/10.1088/1742-6596/551/1/012039} {\bibfield  {journal}
			{\bibinfo  {journal} {J. Phys.: Conf. Ser.}\ }\textbf {\bibinfo {volume}
				{551}},\ \bibinfo {pages} {012039} (\bibinfo {year} {2014})}\BibitemShut
		{NoStop}%
		\bibitem [{\citenamefont {McKenzie}\ \emph {et~al.}(2015)\citenamefont
			{McKenzie}, \citenamefont {Daley}, \citenamefont {Kiefl}, \citenamefont
			{Levy}, \citenamefont {MacFarlane}, \citenamefont {Morris}, \citenamefont
			{Pearson}, \citenamefont {Wang},\ and\ \citenamefont
			{Forrest}}]{2015-McKenzie-SM-11-1755}%
		\BibitemOpen
		\bibfield  {author} {\bibinfo {author} {\bibfnamefont {I.}~\bibnamefont
				{McKenzie}}, \bibinfo {author} {\bibfnamefont {C.~R.}\ \bibnamefont {Daley}},
			\bibinfo {author} {\bibfnamefont {R.~F.}\ \bibnamefont {Kiefl}}, \bibinfo
			{author} {\bibfnamefont {C.~D.~P.}\ \bibnamefont {Levy}}, \bibinfo {author}
			{\bibfnamefont {W.~A.}\ \bibnamefont {MacFarlane}}, \bibinfo {author}
			{\bibfnamefont {G.~D.}\ \bibnamefont {Morris}}, \bibinfo {author}
			{\bibfnamefont {M.~R.}\ \bibnamefont {Pearson}}, \bibinfo {author}
			{\bibfnamefont {D.}~\bibnamefont {Wang}},\ and\ \bibinfo {author}
			{\bibfnamefont {J.~A.}\ \bibnamefont {Forrest}},\ }\bibfield  {title}
		{\enquote {\bibinfo {title} {Enhanced high-frequency molecular dynamics in
					the near-surface region of polystyrene thin films observed with
					{{$\beta$}-NMR}},}\ }\href {https://doi.org/10.1039/C4SM02245A} {\bibfield
			{journal} {\bibinfo  {journal} {Soft Matter}\ }\textbf {\bibinfo {volume}
				{11}},\ \bibinfo {pages} {1755--1761} (\bibinfo {year} {2015})}\BibitemShut
		{NoStop}%
		\bibitem [{\citenamefont {McKenzie}\ \emph {et~al.}(2018)\citenamefont
			{McKenzie}, \citenamefont {Chai}, \citenamefont {Cortie}, \citenamefont
			{Forrest}, \citenamefont {Fujimoto}, \citenamefont {Karner}, \citenamefont
			{Kiefl}, \citenamefont {Levy}, \citenamefont {MacFarlane}, \citenamefont
			{McFadden}, \citenamefont {Morris}, \citenamefont {Pearson},\ and\
			\citenamefont {Zhu}}]{2018-McKenzie-SM-14-7324}%
		\BibitemOpen
		\bibfield  {author} {\bibinfo {author} {\bibfnamefont {I.}~\bibnamefont
				{McKenzie}}, \bibinfo {author} {\bibfnamefont {Y.}~\bibnamefont {Chai}},
			\bibinfo {author} {\bibfnamefont {D.~L.}\ \bibnamefont {Cortie}}, \bibinfo
			{author} {\bibfnamefont {J.~A.}\ \bibnamefont {Forrest}}, \bibinfo {author}
			{\bibfnamefont {D.}~\bibnamefont {Fujimoto}}, \bibinfo {author}
			{\bibfnamefont {V.~L.}\ \bibnamefont {Karner}}, \bibinfo {author}
			{\bibfnamefont {R.~F.}\ \bibnamefont {Kiefl}}, \bibinfo {author}
			{\bibfnamefont {C.~D.~P.}\ \bibnamefont {Levy}}, \bibinfo {author}
			{\bibfnamefont {W.~A.}\ \bibnamefont {MacFarlane}}, \bibinfo {author}
			{\bibfnamefont {R.~M.~L.}\ \bibnamefont {McFadden}}, \bibinfo {author}
			{\bibfnamefont {G.~D.}\ \bibnamefont {Morris}}, \bibinfo {author}
			{\bibfnamefont {M.~R.}\ \bibnamefont {Pearson}},\ and\ \bibinfo {author}
			{\bibfnamefont {S.}~\bibnamefont {Zhu}},\ }\bibfield  {title} {\enquote
			{\bibinfo {title} {Direct measurements of the temperature{,} depth and
					processing dependence of phenyl ring dynamics in polystyrene thin films by
					{$\beta$}-detected {NMR}},}\ }\href {https://doi.org/10.1039/C8SM00812D}
		{\bibfield  {journal} {\bibinfo  {journal} {Soft Matter}\ }\textbf {\bibinfo
				{volume} {14}},\ \bibinfo {pages} {7324--7334} (\bibinfo {year}
			{2018})}\BibitemShut {NoStop}%
		\bibitem [{\citenamefont {Eley}, \citenamefont {Glatz},\ and\ \citenamefont
			{Willa}(2021)}]{Vortex_Eley2021}%
		\BibitemOpen
		\bibfield  {author} {\bibinfo {author} {\bibfnamefont {S.}~\bibnamefont
				{Eley}}, \bibinfo {author} {\bibfnamefont {A.}~\bibnamefont {Glatz}},\ and\
			\bibinfo {author} {\bibfnamefont {R.}~\bibnamefont {Willa}},\ }\bibfield
		{title} {\enquote {\bibinfo {title} {Challenges and transformative
					opportunities in superconductor vortex physics},}\ }\href
		{https://doi.org/10.1063/5.0055611} {\bibfield  {journal} {\bibinfo
				{journal} {J. Appl. Phys.}\ }\textbf {\bibinfo {volume} {130}},\ \bibinfo
			{pages} {050901} (\bibinfo {year} {2021})}\BibitemShut {NoStop}%
		\bibitem [{\citenamefont {Saadaoui}\ \emph
			{et~al.}(2009{\natexlab{a}})\citenamefont {Saadaoui}, \citenamefont
			{MacFarlane}, \citenamefont {Morris}, \citenamefont {Salman}, \citenamefont
			{Chow}, \citenamefont {Fan}, \citenamefont {Hossain}, \citenamefont {Liang},
			\citenamefont {Mansour}, \citenamefont {Parolin}, \citenamefont {Smadella},
			\citenamefont {Song}, \citenamefont {Wang},\ and\ \citenamefont
			{Kiefl}}]{vortex_YCBO_Saadaoui2009}%
		\BibitemOpen
		\bibfield  {author} {\bibinfo {author} {\bibfnamefont {H.}~\bibnamefont
				{Saadaoui}}, \bibinfo {author} {\bibfnamefont {W.}~\bibnamefont
				{MacFarlane}}, \bibinfo {author} {\bibfnamefont {G.}~\bibnamefont {Morris}},
			\bibinfo {author} {\bibfnamefont {Z.}~\bibnamefont {Salman}}, \bibinfo
			{author} {\bibfnamefont {K.}~\bibnamefont {Chow}}, \bibinfo {author}
			{\bibfnamefont {I.}~\bibnamefont {Fan}}, \bibinfo {author} {\bibfnamefont
				{M.}~\bibnamefont {Hossain}}, \bibinfo {author} {\bibfnamefont
				{R.}~\bibnamefont {Liang}}, \bibinfo {author} {\bibfnamefont
				{A.}~\bibnamefont {Mansour}}, \bibinfo {author} {\bibfnamefont
				{T.}~\bibnamefont {Parolin}}, \bibinfo {author} {\bibfnamefont
				{M.}~\bibnamefont {Smadella}}, \bibinfo {author} {\bibfnamefont
				{Q.}~\bibnamefont {Song}}, \bibinfo {author} {\bibfnamefont {D.}~\bibnamefont
				{Wang}},\ and\ \bibinfo {author} {\bibfnamefont {R.}~\bibnamefont {Kiefl}},\
		}\bibfield  {title} {\enquote {\bibinfo {title} {Vortex lattice disorder in
					\ch{YBa2Cu3O_{7-$\delta$}} studied with $\beta$-nmr},}\ }\href
		{https://doi.org/https://doi.org/10.1016/j.physb.2008.11.229} {\bibfield
			{journal} {\bibinfo  {journal} {Physica B}\ }\textbf {\bibinfo {volume}
				{404}},\ \bibinfo {pages} {730--733} (\bibinfo {year}
			{2009}{\natexlab{a}})}\BibitemShut {NoStop}%
		\bibitem [{\citenamefont {Saadaoui}\ \emph
			{et~al.}(2009{\natexlab{b}})\citenamefont {Saadaoui}, \citenamefont
			{MacFarlane}, \citenamefont {Salman}, \citenamefont {Morris}, \citenamefont
			{Song}, \citenamefont {Chow}, \citenamefont {Hossain}, \citenamefont {Levy},
			\citenamefont {Mansour}, \citenamefont {Parolin}, \citenamefont {Pearson},
			\citenamefont {Smadella}, \citenamefont {Wang},\ and\ \citenamefont
			{Kiefl}}]{vortex_YCBO_Saadaoui2009a}%
		\BibitemOpen
		\bibfield  {author} {\bibinfo {author} {\bibfnamefont {H.}~\bibnamefont
				{Saadaoui}}, \bibinfo {author} {\bibfnamefont {W.~A.}\ \bibnamefont
				{MacFarlane}}, \bibinfo {author} {\bibfnamefont {Z.}~\bibnamefont {Salman}},
			\bibinfo {author} {\bibfnamefont {G.~D.}\ \bibnamefont {Morris}}, \bibinfo
			{author} {\bibfnamefont {Q.}~\bibnamefont {Song}}, \bibinfo {author}
			{\bibfnamefont {K.~H.}\ \bibnamefont {Chow}}, \bibinfo {author}
			{\bibfnamefont {M.~D.}\ \bibnamefont {Hossain}}, \bibinfo {author}
			{\bibfnamefont {C.~D.~P.}\ \bibnamefont {Levy}}, \bibinfo {author}
			{\bibfnamefont {A.~I.}\ \bibnamefont {Mansour}}, \bibinfo {author}
			{\bibfnamefont {T.~J.}\ \bibnamefont {Parolin}}, \bibinfo {author}
			{\bibfnamefont {M.~R.}\ \bibnamefont {Pearson}}, \bibinfo {author}
			{\bibfnamefont {M.}~\bibnamefont {Smadella}}, \bibinfo {author}
			{\bibfnamefont {D.}~\bibnamefont {Wang}},\ and\ \bibinfo {author}
			{\bibfnamefont {R.~F.}\ \bibnamefont {Kiefl}},\ }\bibfield  {title} {\enquote
			{\bibinfo {title} {Vortex lattice disorder in \ch{YBa2Cu3O_{7-$\delta$}}
					probed using $\ensuremath{\beta}\text{-NMR}$},}\ }\href
		{https://doi.org/10.1103/PhysRevB.80.224503} {\bibfield  {journal} {\bibinfo
				{journal} {Phys. Rev. B}\ }\textbf {\bibinfo {volume} {80}},\ \bibinfo
			{pages} {224503} (\bibinfo {year} {2009}{\natexlab{b}})}\BibitemShut
		{NoStop}%
		\bibitem [{\citenamefont {Salman}\ \emph {et~al.}(2007)\citenamefont {Salman},
			\citenamefont {Wang}, \citenamefont {Chow}, \citenamefont {Hossain},
			\citenamefont {Kreitzman}, \citenamefont {Keeler}, \citenamefont {Levy},
			\citenamefont {MacFarlane}, \citenamefont {Miller}, \citenamefont {Morris},
			\citenamefont {Parolin}, \citenamefont {Saadaoui}, \citenamefont {Smadella},\
			and\ \citenamefont {Kiefl}}]{vortex_NbSe2_Salman2007}%
		\BibitemOpen
		\bibfield  {author} {\bibinfo {author} {\bibfnamefont {Z.}~\bibnamefont
				{Salman}}, \bibinfo {author} {\bibfnamefont {D.}~\bibnamefont {Wang}},
			\bibinfo {author} {\bibfnamefont {K.~H.}\ \bibnamefont {Chow}}, \bibinfo
			{author} {\bibfnamefont {M.~D.}\ \bibnamefont {Hossain}}, \bibinfo {author}
			{\bibfnamefont {S.~R.}\ \bibnamefont {Kreitzman}}, \bibinfo {author}
			{\bibfnamefont {T.~A.}\ \bibnamefont {Keeler}}, \bibinfo {author}
			{\bibfnamefont {C.~D.~P.}\ \bibnamefont {Levy}}, \bibinfo {author}
			{\bibfnamefont {W.~A.}\ \bibnamefont {MacFarlane}}, \bibinfo {author}
			{\bibfnamefont {R.~I.}\ \bibnamefont {Miller}}, \bibinfo {author}
			{\bibfnamefont {G.~D.}\ \bibnamefont {Morris}}, \bibinfo {author}
			{\bibfnamefont {T.~J.}\ \bibnamefont {Parolin}}, \bibinfo {author}
			{\bibfnamefont {H.}~\bibnamefont {Saadaoui}}, \bibinfo {author}
			{\bibfnamefont {M.}~\bibnamefont {Smadella}},\ and\ \bibinfo {author}
			{\bibfnamefont {R.~F.}\ \bibnamefont {Kiefl}},\ }\bibfield  {title} {\enquote
			{\bibinfo {title} {Magnetic-field effects on the size of vortices below the
					surface of \ch{NbSe2} detected using low energy $\ensuremath{\beta}$-nmr},}\
		}\href {https://doi.org/10.1103/PhysRevLett.98.167001} {\bibfield  {journal}
			{\bibinfo  {journal} {Phys. Rev. Lett.}\ }\textbf {\bibinfo {volume} {98}},\
			\bibinfo {pages} {167001} (\bibinfo {year} {2007})}\BibitemShut {NoStop}%
	\end{thebibliography}
	%
	
\end{document}